\documentclass[twocolumn]{autart}
 
\usepackage{amsmath,amssymb}
\usepackage{mathrsfs}
\usepackage[dvips]{graphicx}
\usepackage{epsfig,psfrag} 
\usepackage{subfigure}
\usepackage{fancyhdr}
\usepackage{color}
\usepackage{verbatim}
\usepackage{natbib}        

\newtheorem{theorem}{Theorem}[section]

\newtheorem{lemma}[theorem]{Lemma}

\newtheorem{definition}{Definition}[section]

\newtheorem{remark}{Remark}[section]

\theoremstyle{definition}
\newtheorem{example}{Example}[section]

\newcommand{\RR}{{\mathbb R}}

\newcommand{\cB}{{\mathcal B}}

\newcommand{\cO}{{\mathcal O}}

\newcommand{\ol}{\overline}

\renewcommand{\Im}{\textup{Im }}
\newcommand{\Ker}{\textup{Ker }}
\newcommand{\aff}{\textup{aff }}

\newcommand{\conv}{\textup{co }}

\newcommand{\spn}{\textup{sp }}

\newcommand{\rank}{\textup{rank}}

\newcommand\xqed[1]{%
  \leavevmode\unskip\penalty9999 \hbox{}\nobreak\hfill
  \quad\hbox{#1}}
\newcommand\demo{\xqed{$\triangleleft$}}

\begin{document}

\begin{frontmatter}
\title{On the Control of Affine Systems with Safety Constraints: Relaxed In-Block Controllability }
\thanks[footnoteinfo]{M.K. Helwa is with the Institute for Aerospace Studies, University of Toronto, and an affiliate member with the Department of Electrical and Computer
Engineering and Centre for Intelligent Machines, McGill University; P. E. Caines is with the Department of Electrical and Computer Engineering and Centre for Intelligent Machines, McGill University (e-mail:
mohamed.helwa@robotics.utias.utoronto.ca, peterc@cim.mcgill.ca).}%
\author{Mohamed K. Helwa},
\author{Peter E. Caines}

\begin{abstract}
We consider affine systems defined on polytopes and study the cases where the systems are not in-block controllable with respect to the given polytopes. That are the cases in which we cannot fully control the affine systems within the interior of a given polytope, representing the intersection of given safety constraints. Instead, we introduce in this paper the notion of relaxed in-block controllability (RIBC), which can be useful for the cases where one can distinguish between soft and hard safety constraints. In particular, we study whether all the states in the interior of a given polytope, formed by the intersection of soft safety constraints, are mutually accessible through the interior of a given bigger polytope, formed by the intersection of hard safety constraints, by applying uniformly bounded control inputs. By exploring the geometry of the problem, we provide necessary conditions for RIBC. We then show when these conditions are also sufficient. Several illustrative examples are also given to clarify the main results. 
\end{abstract}
\end{frontmatter}

\section{Introduction}
The control specifications of modern industrial systems are becoming successively more and more complex; in addition to the traditional requirements of achieving stabilization, tracking and possibly some form of optimized behavior,  contemporary specifications may include safety constraints, human interference, temporal logic statements, and start-up procedures, among others. In the 1990's, hybrid systems were introduced, in part, in order to provide a formalized  system theoretic framework to handle such complex specifications \citep{TEEL,BEMPORAD,BemMor2000}. The reason for this is that hybrid systems combine both continuous dynamics and discrete events, and so they provide a suitable framework for the development of formal verification and synthesis methods for achieving complex specifications.

Recently, special interest has been given to an important class of hybrid systems, namely piecewise affine (PWA) hybrid systems \citep{BemMor2000,HVS06,PWA3,PWA1,HC14,HB15}, since this class of systems has desirable control theoretic properties which hold when simple verifiable conditions are satisfied, can approximate nonlinear systems with arbitrary accuracy \citep{PWA1}, and can be easily identified from experimental data \citep{PWA2}. Interesting applications of PWA control systems can be found in \citep{PWA_app1,PWA_app2,PWA_app3,HHB14}. A PWA hybrid system is typically expressed by a discrete automaton \citep{Hop} such that inside each discrete mode of the automaton, the system is described by affine dynamics defined on a full-dimensional polytope. If the state trajectory of the continuous affine dynamics reaches a prescribed exit facet of the polytope, the PWA hybrid system is transferred to a new discrete mode, in which it evolves according to new affine dynamics in a successive polytope, and so on. Hence, the study of PWA hybrid systems at the continuous level reduces to the study of affine dynamics on full-dimensional polytopes \citep{HVS06}.

In \citep{HC14_2,HC15_TAC}, we introduce the study of the in-block controllability (IBC) of affine systems on polytopes. In particular, for a given affine system and a given full-dimensional polytope $X$, we study whether all the states in the interior of $X$ are mutually accessible through its interior by applying uniformly bounded control inputs. The motivation behind the IBC notion is that it formalizes controllability under safety state constraints. Moreover, we show in \citep{HC14,HC15} that if one constructs a special partition/cover of the state space of PWA hybrid systems/nonlinear systems, in which each region satisfies the IBC property, then one can systematically study controllability properties and build hierarchical control structures of these complex systems. These hierarchical structures are typically used for synthesizing correct-by-design controllers enforcing formal logic specifications \citep{Caines02,Bisim1}. One advantage of the IBC hierarchical structures is that they take into account the fact that different states in a partition/cover region typically need different inputs to be steered to neighborhood regions, possibly over different time horizons. Thus, the IBC hierarchical methods do not require the partition/cover regions to be of very small size, they end up with reasonable number of regions in the partitions/covers, and hence, they may have good potential to be extended to high-dimensional systems. Furthermore, the IBC notion is also useful in the context of optimal control problems \citep{Caines_opt1,Caines_opt2}. In particular, if it is required to find an optimal trajectory connecting two states in the interior of a polytope representing the system's state constraints, then it may be useful to first study IBC to verify that there exists a feasible solution trajectory connecting each pair of states. Then, one may apply the Pontryagin's Minimum Principle to find the optimal path.   

However we have found many examples in which the given affine system is not IBC through the interior of the given polytope $X$, but the mutual accessibility property is achieved if we relax the problem little bit and allow trajectories starting in the interior of $X$ to visit a neighborhood of $X$ in the transient. This is acceptable in many practical scenarios in which one can distinguish between two types of constraints: soft constraints and hard constraints. The soft constraints may form the region of the nominal operating states of the system, while the hard constraints may represent the strict safety constraints which cannot be violated even in the transient period. Hence, it is reasonable to study mutual accessibility of the states in the interior of a polytope $X$, formed by the soft constraints, through the interior of a bigger polytope $X'$ ($X\subset X'$), formed by the hard constraints. This motivates us to introduce and study the relaxed in-block controllability (RIBC) notion in this paper. The study of RIBC is also the first step in extending the hierarchical structures/controllability results in \citep{HC14,HC15} based on the new, relaxed notion.  

The study of controllability is fundamental to modern control systems; as is well-known, for linear systems, algebraic conditions were provided in the 1960's \citep{Kalman}. Restricting our attention here to controllability of linear systems subject to constraints, we next cite \citep{Brammer} and \citep{Sontag}, in which controllability under input constraints was studied. In \citep{Heemels2}, controllability of continuous-time linear systems with input and/or state constraints was studied under the assumption that the transfer matrix of the system is right-invertible. Under the same assumption, \citep{Heemels3} studied null controllability of discrete-time linear systems with input and/or state constraints. The IBC notion formalizes the study of controllability under state constraints. The notion was first introduced in \citep{Caines95} for finite state machines, and was then extended in \citep{Caines98} for continuous nonlinear systems on closed sets and in \citep{Caines02} for automata. In these papers, the IBC notion is used to build hierarchical structures of the systems, but these papers do not study conditions for the IBC property to hold. It is worth mentioning that the IBC concept and its associated between block controllability (BBC) notion in \citep{Caines95,Caines98,Caines02} are entirely different from the bisimulation notion \citep{Bisim1,Bisim2}, also used for constructing system abstractions. In addition to having different axioms, the methods of utilizing these notions to construct the abstractions are different. For instance, while the bisimulation-based methods typically use overapproximation of reachable sets to calculate the abstraction, this is not needed for the IBC hierarchical abstractions. In \citep{HC14_2,HC15_TAC}, we provide three easily checkable necessary and sufficient conditions for IBC to hold for affine systems on polytopes. We then use the results of \citep{HC14_2,HC15_TAC} to study controllability and build hierarchical structures of piecewise affine hybrid systems \citep{HC14}, and to systematically achieve approximate mutual accessibility properties of nonlinear systems under safety constraints \citep{HC15}. In \citep{HC15_2,HS16}, we provide computationally efficient algorithms for building polytopic regions satisfying the IBC property, while In \citep{H15}, we extend the IBC conditions to controlled switched linear systems having both continuous inputs and on/off control switches. We here extend the results concerning the IBC properties developed in \citep{HC14_2,HC15_TAC} to the case where there are soft and hard safety constraints as discussed above. While the IBC notion was used before in \citep{Caines95,Caines98,Caines02}, the RIBC notion is novel. 

After defining RIBC, we explore the geometry of the problem and provide for all the possible geometric cases necessary conditions for RIBC. Then, we show where these conditions are also sufficient. Several illustrative examples are given to clarify the main results. For this study, we exploit some geometric tools used in the study of the controlled invariance problem \citep{Blanchini2,Blanchini} and the reach control problems \citep{HVS04,HVS06,MEB10,H13,HB12,HB13,HB14,HB15,HLB16}. In spite of using similar geometric tools in studying RIBC, our problem is different. Unlike the controlled invariance problem, we do not force all trajectories starting in $X$ to remain in $X$ itself, and we have the additional requirement of achieving mutual accessibility. Also, unlike the reach control problem, in RIBC, we do not try to force the trajectories of the affine system to exit the polytope in finite time through a prescribed facet.

The paper is organized as follows. Section~\ref{sec:back} provides some relevant, mathematical preliminaries.   
In Section~\ref{sec:inblock}, we briefly review IBC.
In Section ~\ref{sec:rel_inblock}, we define RIBC and explore necessary and sufficient conditions for it.
Section \ref{sec:ex} provides examples of the main results. A brief version of this paper appeared in \citep{HC14_3}. Here we include more results and discussions. For instance, we provide here two other cases where the found necessary conditions are also sufficient. We also provide here computational aspects, complete proofs, and a section on examples with simulation results.

{\em Notation}. \ 
Let $K \subset \RR^n$ be a set. 
The closure denotes $\ol{K}$, the interior is denoted $K^{\circ}$, and the boundary is $\partial K$. 
The notation $\dim(K)$ denotes the affine dimension of $K$.
For vectors $x,~y\in \RR^n$, the notation $x.y$ denotes the inner product of the two vectors.
The notation $\left\|x\right\|$ denotes the Euclidean norm of $x$.
For two subspaces $S_1,~S_2$, $S_1+S_2:=\{s_1+s_2~:~s_1\in S_1,~s_2\in S_2\}$.
The notation $\conv(K)$ denotes the convex hull of $K$, while the notation $\conv\{ v_1,v_2,\ldots \}$ denotes the convex hull of a set of points $v_i \in \RR^n$. 
Finally, $B_{\delta}(x)$ denotes the open ball of radius $\delta$ centered at $x$. 

\section{Background}
\label{sec:back}
We provide the relevant geometric background. 
A set $K \subset \RR^n$ is said to be \emph{affine} if for every $x,y\in K$ and every $\alpha\in \RR$, we have $\alpha x+(1-\alpha)y \in K$. Moreover, if $0\in K$, then $K$ is a subspace of $\RR^n$.
A \emph{hyperplane} is an $(n-1)$-dimensional affine set in $\RR^n$, and it divides $\RR^n$ into two open half-spaces.
An \emph{affine hull} of a set $K$, $\aff(K)$, is the smallest affine set containing $K$.
We mean by a dimension of a set $K$ its \emph{affine dimension}, the dimension of $\aff(K)$ \citep{ROCK}. 
A finite collection of vectors $\left\{x_1,\cdots,x_k\right\}$ is called \emph{affinely independent} if the unique solution to $\sum_{i=1}^{k}\alpha_i x_i=0$ and $\sum_{i=1}^{k}\alpha_i =0$ is $\alpha_i=0$ for all $i=1,\cdots,k$. If $\left\{x_1,\cdots,x_k\right\}$ is affinely independent, then these vectors do not lie in a common hyperplane. An \emph{$n$-dimensional simplex} is the convex hull of $(n+1)$ affinely independent points in $\RR^n$, and it generalizes the triangle notion in 2D to arbitrary dimensions.  
   
An $n$-dimensional \emph{polytope} is the convex hull of a finite set of points in $\RR^n$, with dimension $n$ \citep{Brondsted2}. In particular, let $\left\{v_1,\cdots,v_p\right\}$ be a set of points in $\RR^n$, where $p>n$, and suppose that $\left\{v_1,\cdots,v_p\right\}$ contains (at least) $n+1$ affinely independent points. We denote the $n$-dimensional polytope generated by $\left\{v_1,\cdots,v_p\right\}$ by $X:=\conv\left\{v_1,\cdots,v_p\right\}$. Note that an $n$-dimensional simplex is a special case of $X$ with $p=n+1$.
A \emph{face} of $X$ is any intersection of $X$ with a closed half-space such that none of the interior points of $X$ lie on the boundary of the half-space. 
According to this definition, the polytope $X$ and the empty set are considered as trivial faces, and we call all other faces \emph{proper faces}.
A \emph{facet} of $X$ is an $(n-1)$-dimensional face of $X$. We denote the facets of $X$ by $F_1,\cdots,F_r$, and we use $h_i$ to denote the unit normal vector to $F_i$ pointing outside $X$. 
An $n$-dimensional polytope $X$ is \emph{simplicial} if all its facets are simplices.
It is clear that any two-dimensional convex polytope is simplicial. For higher dimensions, convex compact sets can be approximated by simplicial polytopes with arbitrary accuracy \citep{simplicial1,simplicial2}.

We conclude this section by reviewing the definition of the Bouligand tangent cone of a closed set \citep{Clarke}. Let $S \subset \RR^n$ be a closed set. We define the distance function $d_{S}(x):=\inf \left\{\left\|x-y\right\|~:~y\in S\right\}$.
The \emph{Bouligand tangent cone} (or simply tangent cone) to $S$ at $x\in S$, denoted $T_{S}(x)$, is defined by
\[
T_{S}(x):=\left\{v\in \RR^n~:~\liminf_{t\rightarrow 0^+}\frac{d_{S}(x+tv)}{t}=0\right\}.
\]
If $S$ is convex, so is $T_{S}(x)$ \citep{Blanchini}.
 
\section{In-Block Controllability}
\label{sec:inblock}
In this section we briefly review the in-block controllability (IBC) \citep{HC14_2}, and then we provide a motivating example for defining the relaxed in-block controllability (RIBC) in the next section.
Consider the affine control system:
\begin{equation}
\label{eq:thesystem}
\dot x =A x + Bu + a \,,~~~x\in\RR^n,
\end{equation}
where $A \in \mathbb{R}^{n \times n}$, $a \in \mathbb{R}^n$, 
$B \in \mathbb{R}^{n \times m}$, and $\rank(B)=m$.
In this paper, we assume that the control input $u:[0,\infty)\rightarrow \RR^m$ is measurable and bounded on any compact time interval to ensure the existence and uniqueness of the solutions of \eqref{eq:thesystem} \citep{Filippov}.
Let $\cB:=\Im(B)$, the image of $B$, and let $\phi(x_0,t,u)$ be the trajectory of \eqref{eq:thesystem}, under a control input $u$, with initial condition $x_0 \in X$, and evaluated at time instant $t$. We review the IBC definition (after \citep{Caines98}).
\begin{definition}[In-Block Controllability (IBC)] 
\label{prob0}
Consider the affine control system \eqref{eq:thesystem} on an $n$-dimensional polytope $X$. We say that \eqref{eq:thesystem} is in-block controllable (IBC) w.r.t. $X$ if there exists $M>0$ such that for all $x,~y\in X^{\circ}$, there exist $T\geq 0$ and a control input $u$ defined on $[0,T]$ such that (i) $\left\|u(t)\right\|\leq M$ and $\phi(x,t,u)\in X^{\circ}$ for all $t\in [0,T]$, and (ii) $\phi(x,T,u)=y$.   
\end{definition}

In \citep{HC14_2}, it is shown that studying IBC of an affine system on a given polytope is equivalent to studying IBC of a linear system 
\begin{equation}
\label{eq:thesystem2}
\dot{x}=Ax+Bu
\end{equation}
on a new polytope $X$ satisfying $0\in X^{\circ}$ (without loss of generality (w.l.o.g.) we use the same notation $X$ for the new polytope). We review the main result of \citep{HC14_2}. To that end, let $J:=\left\{1,\cdots,r\right\}$ and $J(x):=\left\{j\in J~:~x\in F_j\right\}$. That is, $J$ is the set of indices of the facets of $X$, while $J(x)$ is the set of indices of the facets of $X$ in which $x$ is a point. We define the closed, tangent cone to the polytope $X$ at $x$ as $C(x):=\left\{y\in \RR^n~:~h_j\cdot y\leq 0,~j \in J(x)\right\}$, where $h_j$ is the unit normal vector to $F_j$ pointing outside $X$.  
\begin{theorem}[\citep{HC14_2}]
\label{thm:main_IBC}
Consider the system \eqref{eq:thesystem2} defined on an $n$-dimensional simplicial polytope $X$ satisfying $0\in X^{\circ}$. The system \eqref{eq:thesystem2} is IBC w.r.t. $X$ if and only if
\begin{itemize}
\item [(i)] $(A,B)$ is controllable.
\item [(ii)] The so-called invariance conditions of $X$ are solvable (That is, for each vertex $v\in \partial X$, there exists $u\in \RR^m$ such that $Av+Bu\in C(v)$).
\item [(iii)] The so-called backward invariance conditions of $X$ are solvable (That is, for each vertex $v\in \partial X$, there exists $u\in \RR^m$ such that $-Av-Bu\in C(v)$).
\end{itemize}
\end{theorem}
In \citep{HC14_2}, it is also shown that conditions (i)-(iii) of Theorem \ref{thm:main_IBC} are necessary for non-simplicial polytopes. Notice that checking solvability of the invariance conditions (or the backward invariance conditions) can be simply carried out by solving a linear programming (LP) problem at each vertex of $X$. Using a straightforward convexity argument, it can be shown that solvability of the invariance conditions (or the backward invariance conditions) at the vertices implies that they are solvable at all the boundary points of $X$ \citep{HVS04}. 

Nevertheless, the IBC notion is a restrictive one. In particular, we have found in many examples that achieving the conditions (ii), (iii) of Theorem \ref{thm:main_IBC} simultaneously is quite difficult. This motivates us to propose a relaxation of IBC in this paper. In particular, suppose that condition (ii) or (iii) of Theorem \ref{thm:main_IBC} is not achieved. The question arises of whether it is still possible to achieve mutual accessibility of points in $X^{\circ}$ through a bigger polytope. 
To that end, let $\lambda>1$, and define 
$\lambda X:=\left\{x\in \RR^n~:~x=\lambda y,y\in X\right\}$, a $\lambda$-scaled version of $X$.
We start with the following technical result that shows if condition (i) of Theorem \ref{thm:main_IBC} is achieved, then as expected there is always a bigger polytope through which the points of $X^{\circ}$ are mutually accessible.
\begin{lemma}
\label{lem:tec1}
Consider the system \eqref{eq:thesystem2} and an $n$-dimensional convex compact set $X$ satisfying $0\in X^{\circ}$. If $(A,B)$ is controllable, then there exist $M>0$ and $\lambda>1$ such that for all $x,y\in X^{\circ}$, there exist $T\geq 0$ and a control input $u$ defined on $[0,T]$ such that $\left\|u(t)\right\|\leq M$ and $\phi(x,t,u)\in (\lambda X)^{\circ}$ for all $t\in [0,T]$, and $\phi(x,T,u)=y$.  
\end{lemma}
\begin{pf}
See the appendix.
\end{pf}

The following example shows that not every $\lambda>1$ works, and so a careful study is needed to investigate when a given $\lambda$ has the desired properties.
\begin{example}
\label{ex:mot2}
Consider the system
\begin{equation}
\label{sys:ex2}
\dot{x} =
\left[ 
\begin{array}{cc}
0 & 1  \\
0 & 0  
\end{array} 
\right] x +
\left[ 
\begin{array}{rr}
1\\1\end{array} 
\right] u \,
\end{equation} 
\begin{figure}[t]
\begin{center}
\includegraphics[scale=.22, trim = 10mm 100mm 10mm 16mm]{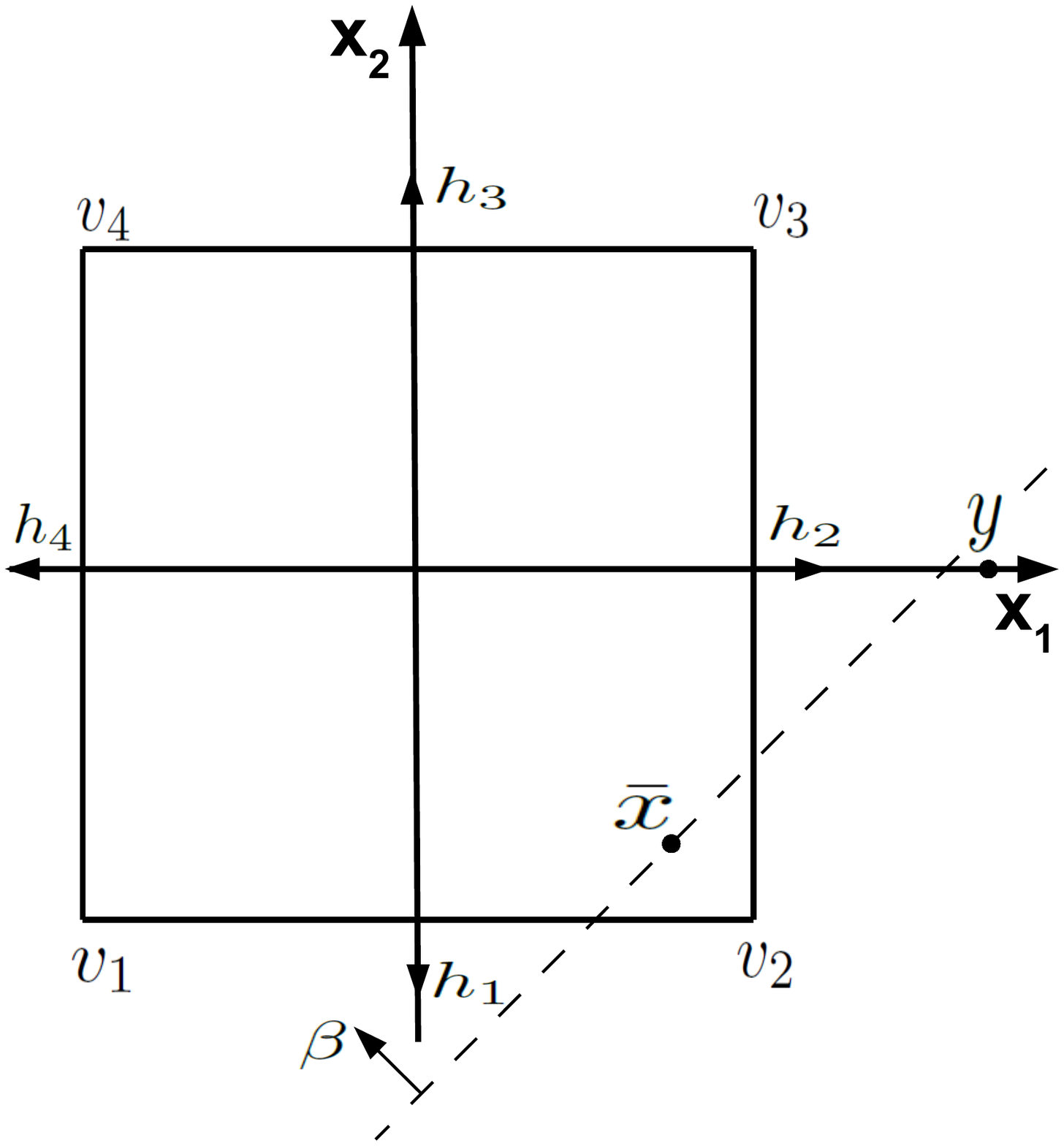} 
\end{center}
\caption{$X$ for Example \ref{ex:mot2}}
\label{fig:ex2}
\end{figure}
and a polytope $X:=\conv\left\{v_1,\cdots,v_4\right\}$ shown in Figure \ref{fig:ex2} where $v_1=(-1,-1)$, $v_2=(1,-1)$, $v_3=(1,1)$, and $v_4=(-1,1)$. 
First, we check whether \eqref{sys:ex2} is IBC w.r.t. $X$. It is easy to verify that $(A,B)$ is controllable. Next, we check solvability of the invariance conditions and the backward invariance conditions of $X$. At $v_2$, we have $-Av_2-Bu_2=(1-u_2,-u_2)$. Solvability of the backward invariance conditions of $X$ at $v_2$ requires the existence of $u_2\in \RR$ such that $-Av_2-Bu_2\in C(v_2)$. This yields $h_1\cdot(-Av_2-Bu_2)\leq 0$ and $h_2\cdot(-Av_2-Bu_2)\leq 0$, where $h_1=(0,-1)$ and $h_2=(1,0)$ as shown in Figure \ref{fig:ex2}. That is, $u_2\leq 0$ and $1-u_2\leq 0$ respectively. Therefore, there does not exist $u_2\in \RR$ that satisfies the backward invariance conditions of $X$ at $v_2$. From Theorem \ref{thm:main_IBC}, the system \eqref{sys:ex2} is not IBC w.r.t. $X$.

Next, we investigate whether in this example $\lambda$, defined in Lemma \ref{lem:tec1}, can be selected arbitrarily close to $1$. Let $\beta:=(-1,1)$. We make several observations. First, we have $\beta \in \Ker(B^T)$, the perpendicular subspace to $\cB$, i.e. $\beta \cdot B=0$. Then, we have $\beta \cdot (-Ax-Bu)=x_2$, which is negative for any $x\in \RR^2$ having $x_2<0$ whatever $u$ is selected. Let $\bar{x}:=(0.9,-0.9)$. It can be easily verified that $\bar{x}\in X^{\circ}$ and $\beta\cdot \bar{x}=-1.8$. Since $(A,B)$ is controllable, we know from Lemma \ref{lem:tec1} there exists $\lambda>1$ such that all the points in $X^{\circ}$ are mutually accessible through $(\lambda X)^{\circ}$ by applying uniformly bounded control inputs. In particular, there exists a state trajectory connecting the origin to $\bar{x}$ in finite time through $(\lambda X)^{\circ}$. Equivalently, there is a state trajectory of the backward dynamics $\dot{x}=-Ax-Bu$ that connects $\bar{x}$ to the origin in finite time through $(\lambda X)^{\circ}$. Since $\beta\cdot 0>\beta\cdot \bar{x}$ and $\beta \cdot (-Ax-Bu)<0$ when $x_2$ is negative, it follows that the state trajectory of the backward dynamics starting at $\bar{x}$ must cross the $x_1$-axis before reaching $0$. Let the point $y$ denote the intersection of the state trajectory with the $x_1$-axis. Since the $\beta$-component of the state trajectory of the backward dynamics is decreasing as long as $x_2$ is still negative, then $\beta \cdot y<\beta \cdot \bar{x}=-1.8$. Since $y$ has a zero $x_2$-component, then the $x_1$-component of $y$ must be greater than $1.8$. Recall that $y\in (\lambda X)^{\circ}$, and so clearly $\lambda>1.8$. We conclude $\lambda$ cannot be selected arbitrarily close to $1$ in this example. \demo            
\end{example}
Inspired by the above example, we define in the next section the relaxed in-block controllability (RIBC). In particular, to tackle this problem, we assume that we have in hand a given polytope $X'$ satisfying $X\subset X'$ and study whether all the states in $X^{\circ}$ are mutually accessible through $X'^{\circ}$ using uniformly bounded inputs. 
       
\section{Relaxed In-Block Controllability}
\label{sec:rel_inblock}
Inspired by our discussion in the previous section, we define the relaxed in-block controllability (RIBC) as follows.
\begin{definition}[Relaxed In-Block Controllability]
\label{prob0_rel}
Consider the affine control system \eqref{eq:thesystem} and $n$-dimensional polytopes $X,~X'$ such that $X\subset X'$. We say that \eqref{eq:thesystem} is relaxed in-block controllable (RIBC) w.r.t. $X$ through $X'$ if there exists $M>0$ such that for all $x,~y\in X^{\circ}$, there exist $T\geq 0$ and a control input $u$ defined on $[0,T]$ such that (i) $\left\|u(t)\right\|\leq M$ and $\phi(x,t,u)\in X'^{\circ}$ for all $t\in [0,T]$, and (ii) $\phi(x,T,u)=y$.  
\end{definition}

\begin{figure}[t] 
\begin{center}
\subfigure[]
{\includegraphics[trim = 10mm 125mm 10mm 40mm,clip,width = 0.31\linewidth]{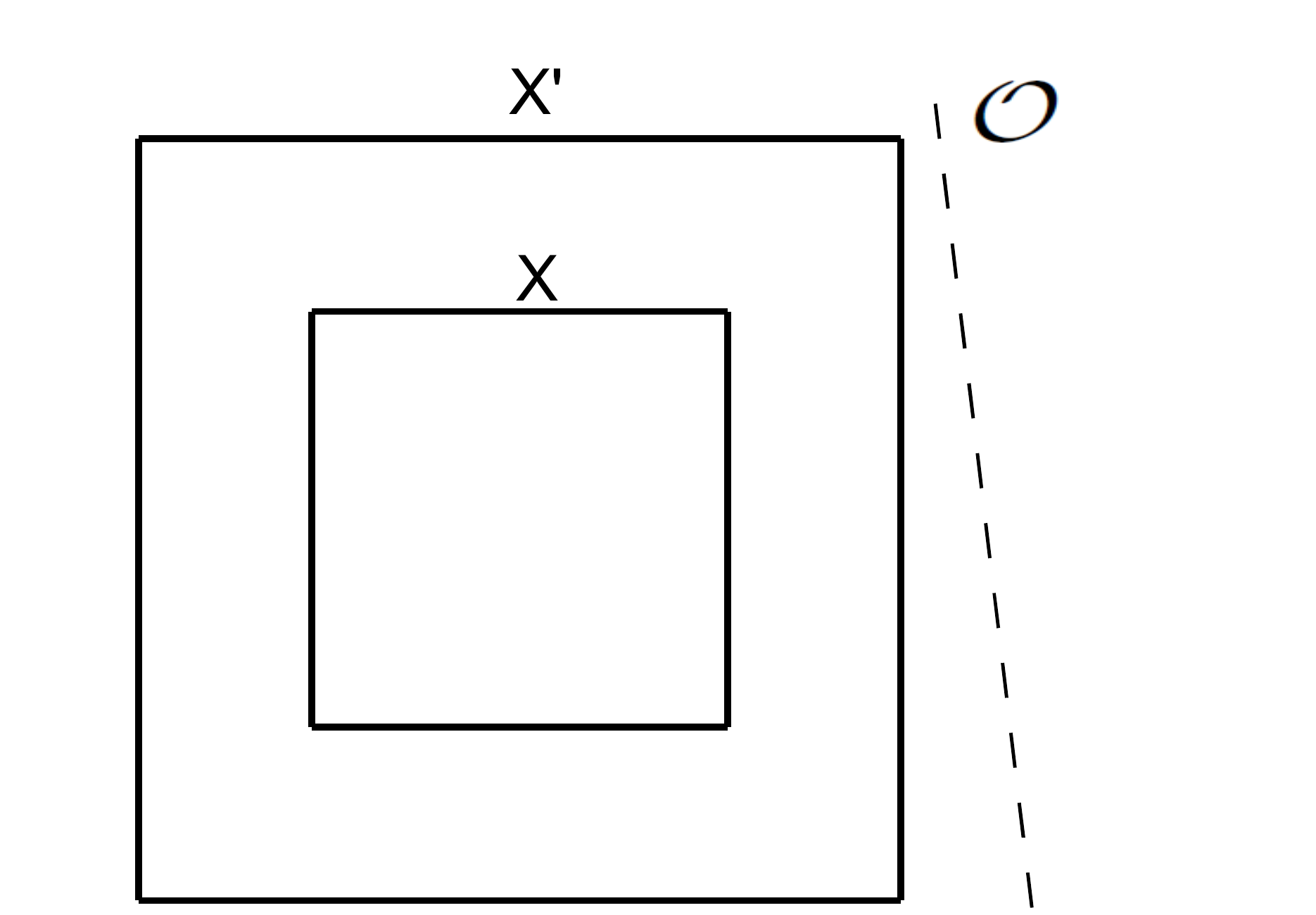}}
\subfigure[]
{\includegraphics[trim = 10mm 125mm 10mm 40mm,clip,width = 0.31\linewidth]{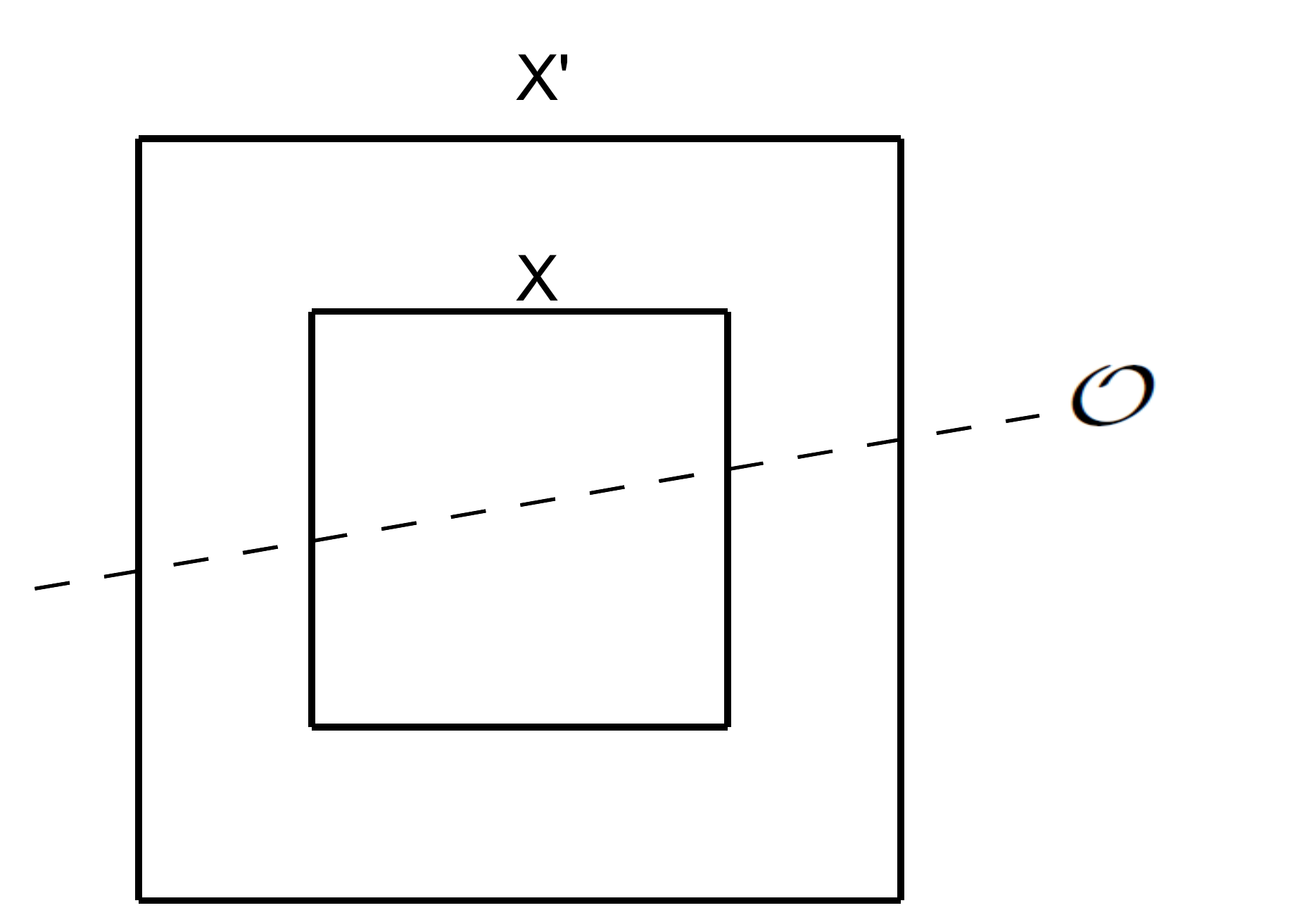}}
\subfigure[]
{\includegraphics[trim = 10mm 125mm 10mm 40mm,clip,width = 0.31\linewidth]{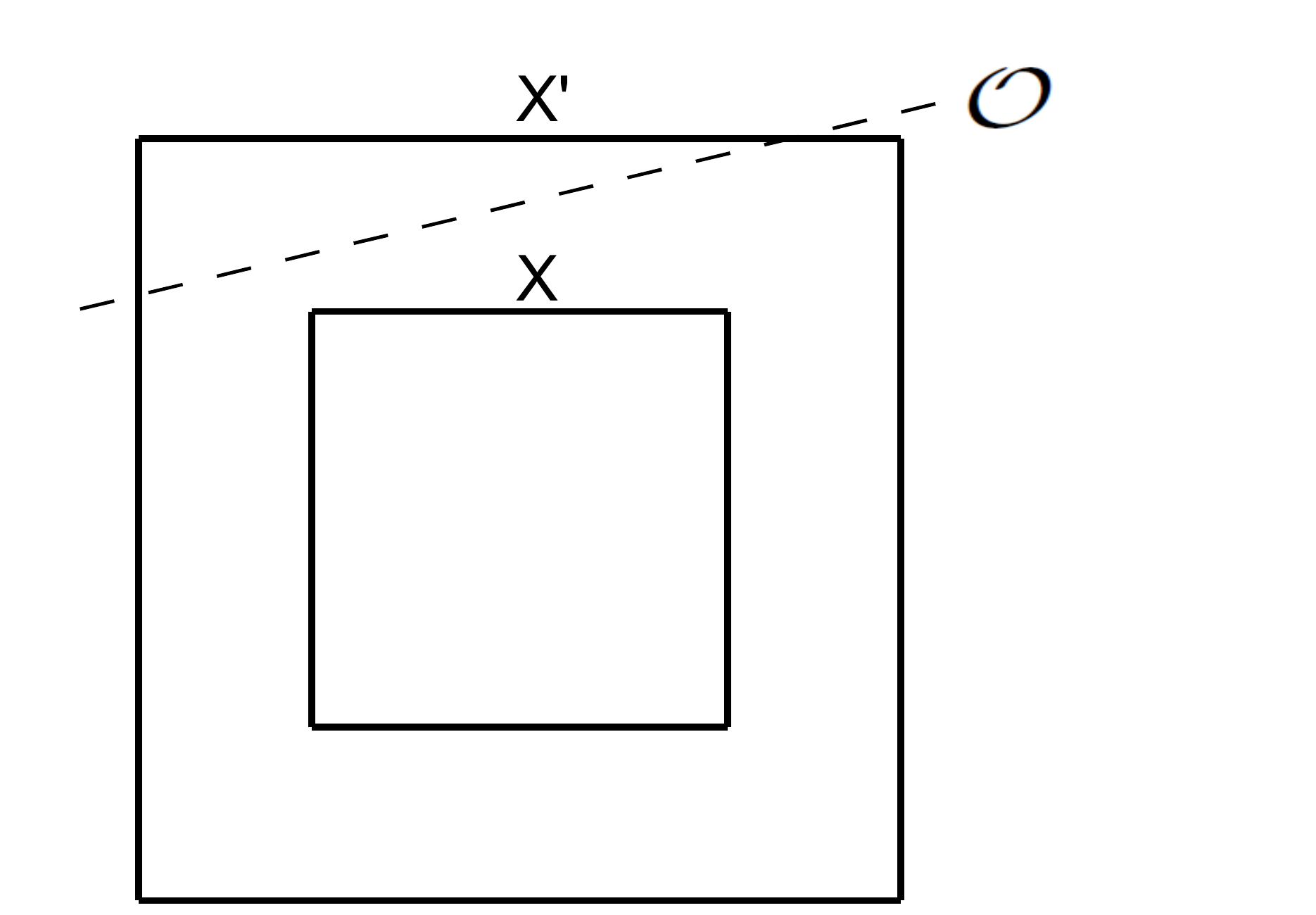}}
\caption{Three geometric cases of the location of $\cO$ w.r.t. $X,~X'$}
\label{fig:O}
\end{center}
\end{figure}

Following \citep{HC14_2}, we use a geometric approach in studying RIBC in this paper. In particular, we define the set of possible equilibria of \eqref{eq:thesystem} as follows \citep{MEB10}:
\begin{equation}
\label{BASICS:O}
\cO := \{ ~ x \in \RR^n ~:~ A x + a \in \cB ~ \} \,.
\end{equation}
The vector field of the system \eqref{eq:thesystem} can vanish at any 
$x \in \cO$ for a proper selection of $u \in \RR^m$. In fact, $\cO$ is the set of all possible
equilibrium points of \eqref{eq:thesystem}, i.e. if $x_0$ is an equilibrium of \eqref{eq:thesystem} under feedback control, then $x_0 \in \cO$. It can be shown that $\cO$ is closed and affine. Also, if $\cO\neq \emptyset$, then $\cO$ has dimension $m \leq \kappa\leq n$ \citep{HB13}. Similarly to the case of IBC \citep{HC14_2}, the location of the set $\cO$ with respect to $X,~X'$ affects RIBC. Thus, in order to simplify our study of RIBC, we classify our study of RIBC into three geometric cases based on the location of $\cO$ with respect to $X,~X'$, namely (i) Case (A): $X'^{\circ}\cap \cO=\emptyset$, (ii) Case (B): $X^{\circ}\cap \cO\neq\emptyset$, and (iii) Case (C): $X^{\circ}\cap \cO=\emptyset$ but $X'^{\circ}\cap \cO\neq \emptyset$ as shown in Figure \ref{fig:O}.
\subsection{$X'^{\circ}\cap \cO=\emptyset$}
This geometric situation is shown in Figure \ref{fig:O}(a). For this case, the following result shows that the affine system \eqref{eq:thesystem} is not RIBC w.r.t. $X$ through $X'$.
\begin{theorem}
\label{thm:caseA}
Consider the affine control system \eqref{eq:thesystem} and $n$-dimensional polytopes $X,~X'$ such that $X\subset X'$. If $X'^{\circ}\cap \cO=\emptyset$, then the system \eqref{eq:thesystem} is not RIBC w.r.t. $X$ through $X'$. 
\end{theorem}
\begin{pf}
The proof is similar to the proof of Theorem 3.1 of \citep{HC14_2}. We provide a brief sketch of it in the appendix for completeness of this draft version of the paper.  
\end{pf}
An illustrative example that clarifies the proof of Theorem \ref{thm:caseA} is provided in Section \ref{sec:ex} (See Example \ref{ex2_sec}).
\subsection{$X^{\circ}\cap \cO\neq\emptyset$}
This geometric situation is shown in Figure \ref{fig:O}(b). For this case, select $\bar{x}\in X^{\circ}\cap \cO$, and let $\bar{u}\in \RR^m$ be such that $A\bar{x}+B\bar{u}+a=0$. This is always possible since $\bar{x}\in \cO$. Define $\tilde{x}=x-\bar{x}$ and $\tilde{u}=u-\bar{u}$. The dynamics in the new coordinates are $\dot{\tilde{x}}=A\tilde{x}+B\tilde{u}$. Therefore, for this geometric case, we can assume w.l.o.g. that we study conditions for RIBC of the linear system \eqref{eq:thesystem2} w.r.t. a polytope $X$ through a polytope $X'$ such that $0\in X^{\circ} \subset X'^{\circ}$. 

\begin{theorem}
\label{thm:caseB_1}
Consider the system \eqref{eq:thesystem2} and $n$-dimensional polytopes $X,~X'$ such that $0\in X^{\circ}\subset X'^{\circ}$. If \eqref{eq:thesystem2} is RIBC w.r.t. $X$ through $X'$, then $(A,B)$ is controllable.  
\end{theorem}
\begin{pf}
The proof is the same as the proof of Theorem 4.1 of \citep{HC14_2}. A sketch of the proof is provided in the appendix for completeness of this draft version.
\end{pf}
Next, we present a second necessary condition for RIBC in this case. To that end, for a convex compact set $X$, we say that \emph{the invariance conditions of $X$ are solvable} (w.r.t. the system \eqref{eq:thesystem2}) if for each $x\in \partial X$, there exists $u\in \RR^m$ such that $Ax+Bu\in T_{X}(x)$, the Bouligand tangent cone to $X$ at $x$ (Notice that if $X$ is an $n$-dimensional polytope, then $T_{X}(x)=C(x)$, which is defined in the previous section).
\begin{theorem}
\label{thm:caseB_2}
Consider the system \eqref{eq:thesystem2} and $n$-dimensional polytopes $X,~X'$ such that $0\in X^{\circ}\subset X'^{\circ}$. If \eqref{eq:thesystem2} is RIBC w.r.t. $X$ through $X'$, then there exists an $n$-dimensional compact convex set $X_1$ such that $X\subseteq X_1 \subseteq X'$ and the invariance conditions of $X_1$ are solvable. 
\end{theorem}
\begin{pf}
By assumption, for each $x_0\in X^{\circ}$, there exists a state trajectory $\phi(x_0,t,u)$ that connects $x_0$ to $0\in X^{\circ}$ in finite time $T_{x_0}$ through $X'^{\circ}$. Therefore, we have
$X^{\circ} \subseteq \left\{\phi(x_0,t,u)~:~t\in [0,T_{x_0}],~x_0\in X^{\circ}\right\}\subseteq X'^{\circ}$.
Since $X'^{\circ}$ is convex, then we have
\[
X^{\circ} \subseteq C:=\conv \left\{\phi(x_0,t,u)~:~t\in [0,T_{x_0}],~x_0\in X^{\circ}\right\}\subseteq X'^{\circ}.
\]
Now we show for each $x\in C$, there exist $T\geq 0$ and a control input $u$ defined on $[0,T]$ such that $\left\|u(t)\right\|\leq M$ (where $M$ is as defined in Definition \ref{prob0_rel}), $\phi(x,t,u)\in C$ for all $t\in [0,T]$, and $\phi(x,T,u)=0$. This is obvious for points in $\left\{\phi(x_0,t,u)~:~t\in [0,T_{x_0}],~x_0\in X^{\circ}\right\}$. For $x_i\in \left\{\phi(x_0,t,u)~:~t\in [0,T_{x_0}],~x_0\in X^{\circ}\right\}$, let $T_{x_i}$ and $u_{x_i}$ denote the time of reaching the origin and the uniformly bounded control input achieving that, respectively.
Then, consider an arbitrary point $x'\in C$, $x'\notin \left\{\phi(x_0,t,u)~:~t\in [0,T_{x_0}],~x_0\in X^{\circ}\right\}$. By the definition of the convex hull, there exist points $x_1,\cdots,x_l\in \left\{\phi(x_0,t,u)~:~t\in [0,T_{x_0}],~x_0\in X^{\circ}\right\}$ such that $x'=\sum_{i=1}^{l}\alpha_ix_i$, where $\alpha_i\geq 0$ and $\sum_{i=1}^{l}\alpha_i=1$. By Caratheodory's Theorem (Theorem 17.1 of \citep{ROCK}), we know that w.l.o.g. $l<\infty$. Let $T:=\max\left\{T_{x_1},\cdots,T_{x_l}\right\}$. For each $x_i$, define the control sequence $\overline{u}_{x_i}(t)$ as follows: $\overline{u}_{x_i}(t)=u_{x_i}(t)$ for all $t\in [0,T_{x_i})$, and $\overline{u}_{x_i}(t)=0$ for all $t\in [T_{x_i},T]$. The control sequence $\overline{u}_{x_i}(t)$ drives the system from $x_i$ to $0$ in finite time $T_{x_i}$ through $C$, then keeps the system at $0$ till time $T>0$. Now starting at $x'$, apply the control sequence $u'(t)=\sum_{i=1}^{l}\alpha_i\overline{u}_{x_i}(t)$, for all $t\in [0,T]$. Since \eqref{eq:thesystem2} is linear, it can be shown by a standard argument that the control inputs $u'$ drive the system from $x'$ to $0$ in finite time $T_{x'}\leq T$ through $C$. Also, since $\left\|\overline{u}_{x_i}(t)\right\|\leq M$ for all $t\in [0,T]$, we have $\left\|u'(t)\right\|\leq M$ for all $t\in [0,T]$.  

Next, let $X_1:=\overline{C}$. Clearly, $X_1$ is convex and closed. Then, since $C\subseteq  X'^{\circ}$, $X_1 \subseteq X'$, so $X_1$ is bounded, hence compact. Also, since $X \subseteq X_1$, $X_1$ is an $n$-dimensional set. It remains to show that the invariance conditions of $X_1$ are solvable. Aided with the property proved in the previous paragraph, this can be shown from the proof of Theorem 4.3 and Remark 4.2 of \citep{HC14_2}. 
\end{pf}
\begin{remark}
\label{rem:inv}
If a polytope is given, then checking its invariance conditions is reduced to solving LP problems at the vertices of the given polytope. However, finding a polytope such that its invariance conditions are solvable requires, in general, solving bilinear matrix inequalities (BMIs) \citep{Blanchini2}, the solving of which is NP-hard \citep{NP}. Since the construction of the polytopes satisfying the invariance conditions is carried out offline, available software packages for solving BMIs such as PENBMI \citep{PENBMI} can be utilized. There is also a wide literature on constructing polytopic invariant sets \citep{Blanchini2}. Instead, note that for \eqref{eq:thesystem2}, $\cO=\{x\in\RR^n~:~Ax\in \cB\}$ is a subspace. If $\cO + \cB=\RR^n$, then one can systematically construct an invariant polytope $X_1$ containing the given polytope $X$, without the need for solving BMIs. Assume w.l.o.g. that the  the invariance conditions of $X$ are not solvable at the vertices $v_1,\cdots,v_L$. For a vertex $v_i$ of $X$, $i\in\{1,\cdots,L\}$, express it as $v_i=o_i+b_i$, where $o_i\in \cO$ and $b_i\in \cB$, which is always possible since $\cO + \cB=\RR^n$. Let $\bar{o}_i:=\alpha o_i$, for some $\alpha>1$. Since $\cO$ is affine, $\bar{o}_i\in \cO$. Define $X_1:=\conv\{v_1,\cdots,v_p,\bar{o}_1,\cdots,\bar{o}_L\}$, where $\{v_1,\cdots,v_p\}$ are the vertices of $X$. By construction, $X\subset X_1$, and it can be shown that $o_i\in X_1^{\circ}$. Now we show the invariance conditions of $X_1$ are solvable. At the vertices $v_{L+1},\cdots,v_p$, select control inputs satisfying the invariance conditions of $X$, and at $\bar{o}_i\in \cO$, select $\bar{u}_i$ such that $A\bar{o}_i+B\bar{u}_i=0$, which achieves the invariance conditions at $\bar{o}_i$. Finally, at $v_i$, $i\in\{1,\cdots,L\}$, since $v_i-b_i=o_i$, $-b_i$ points toward $o_i$, and so points inside the interior of the tangent cone to $X_1$ at $v_i$. Thus, at $v_i$, $i\in\{1,\cdots,L\}$, one can always select the control input in the direction of $-b_i\in \cB$ sufficiently large to push the vector field at $v_i$ close enough to $-b_i$, so that it lies in the interior of the tangent cone to $X_1$ at $v_i$. Instead of the invariant polytopes, one may search for ellipsoidal invariant sets, which are popular candidates since they can be found from the Lyapunov equation or the Ricatti equation \citep{Blanchini2}. For instance, one can solve the linear matrix inequalities (LMIs): $Q>0$ and $QA^T+AQ+Y^TB^T+BY<0$. Then, the Lyapunov function is $V:=x^TPx$, where $P=Q^{-1}>0$, and the corresponding stabilizing feedback is $u=Kx$, where $K:=YP$ \citep{Blanchini2}. An invariant ellipsoidal set containing $X$ is $X_1:=\{x\in \RR^n~:~x^TPx\leq c\}$, where $c\geq \max_{i\in\{1,\cdots,p\}}v_i^TPv_i$. Examples are provided in Section \ref{sec:ex}.       
\end{remark} 

We present a third necessary condition for RIBC in this case. For a convex compact set $X$, we say that \emph{the backward invariance conditions of $X$ are solvable} if for each $x\in \partial X$, there exists $u\in \RR^m$ such that $-Ax-Bu\in T_{X}(x)$. 
\begin{theorem}
\label{thm:caseB_3}
Consider the system \eqref{eq:thesystem2} and $n$-dimensional polytopes $X,~X'$ such that $0\in X^{\circ}\subset X'^{\circ}$. If \eqref{eq:thesystem2} is RIBC w.r.t. $X$ through $X'$, then there exists an $n$-dimensional compact convex set $X_2$ such that $X\subseteq X_2 \subseteq X'$ and the backward invariance conditions of $X_2$ are solvable. 
\end{theorem}
\begin{pf}
By assumption, for each $x\in X^{\circ}$, there exists a state trajectory of \eqref{eq:thesystem2} that connects $0\in X^{\circ}$ to $x$ in finite time through $X'^{\circ}$. Equivalently, for each $x\in X^{\circ}$, there exists a state trajectory of the backward system $\dot{x}=-Ax-Bu$ that connects $x$ to $0$ in finite time through $X'^{\circ}$. The rest of the proof is the same as the proof of Theorem \ref{thm:caseB_2} (for the trajectory of the backward system).
\end{pf}
Now we present the main result for this geometric case.
\begin{theorem}
\label{thm:caseB_main}
Consider the system \eqref{eq:thesystem2} and $n$-dimensional polytopes $X,~X'$ such that $0\in X^{\circ}\subset X'^{\circ}$. If \eqref{eq:thesystem2} is RIBC w.r.t. $X$ through $X'$, then the following conditions hold:
\begin{itemize}
\item [(i)] $(A,B)$ is controllable.
\item [(ii)] There exists an $n$-dimensional compact convex set $X_1$ such that $X\subseteq X_1 \subseteq X'$ and the invariance conditions of $X_1$ are solvable.
\item [(iii)] There exists an $n$-dimensional compact convex set $X_2$ such that $X\subseteq X_2 \subseteq X'$ and the backward invariance conditions of $X_2$ are solvable.
\end{itemize}
Moreover, if $X_1,~X_2$ are simplicial polytopes, then the conditions (i)-(iii) are also sufficient for RIBC.
\end{theorem}
\begin{pf}
The fact that conditions (i)-(iii) are necessary follows directly from Theorems \ref{thm:caseB_1}, \ref{thm:caseB_2}, and \ref{thm:caseB_3} respectively. Now we prove the sufficiency of (i)-(iii). Since $X_1$ is a simplicial polytope satisfying $0\in X_1^{\circ}$ by assumption, the conditions (i), (ii) imply by Theorem 5.9 of \citep{HC14_2} that there exists $M_1>0$ such that for each $x_0\in X_1^{\circ}$, there exist $T\geq 0$ and a control input $u$ defined on $[0,T]$ such that $\left\|u(t)\right\|\leq M_1$ and $\phi(x_0,t,u)\in X_1^{\circ}$ for all $t\in [0,T]$, and $\phi(x_0,T,u)=0$. In particular, all the points in $X^{\circ}\subseteq X_1^{\circ}$ can reach $0$ in finite time through $X_1^{\circ}\subseteq X'^{\circ}$ by applying uniformly bounded control inputs. Then, since $X_2$ is a simplicial polytope satisfying $0\in X_2^{\circ}$ by assumption, the conditions (i), (iii) imply by Theorem 5.10 of \citep{HC14_2} that there exists $M_2>0$ such that for each $x_f\in X_2^{\circ}$, there exist $T\geq 0$ and a control input $u$ defined on $[0,T]$ such that $\left\|u(t)\right\|\leq M_2$ and $\phi(0,t,u)\in X_2^{\circ}$ for all $t\in [0,T]$, and $\phi(0,T,u)=x_f$. In particular, all the points in $X^{\circ}\subseteq X_2^{\circ}$ are accessible from $0$ in finite time through $X_2^{\circ}\subseteq X'^{\circ}$ by applying uniformly bounded control inputs.   
\end{pf}
\begin{remark}
Notice that the sets $X_1,~X_2$ in conditions (ii), (iii) of Theorem \ref{thm:caseB_main} are not necessarily the same set, and this relaxes conditions (ii), (iii) of Theorem \ref{thm:main_IBC} which require solvability of the invariance and the backward invariance conditions for the same set $X$. 
\end{remark} 
\begin{remark}
\label{rem:suff}
Conditions (i)-(iii) of Theorem \ref{thm:caseB_main} are also sufficient under other conditions rather than the condition that $X_1$ and $X_2$ are simplicial polytopes. One of these conditions is that $X_1$ and $X_2$ are polytopes, and the strict invariance conditions (strict backward invariance conditions) are solvable for $X_1$ ($X_2$), respectively. That is to say for each $x\in \partial X_1$ ($x\in \partial X_2$), there exists $u\in \RR^m$ such that $Ax+Bu\in C_1^{\circ}(x)$ ($-Ax-Bu\in C_2^{\circ}(x)$), where $C_1^{\circ}(x)$ ($C_2^{\circ}$(x)) is the interior of the tangent cone to the polytope $X_1$ ($X_2$) at $x$, respectively. The idea of the proof in this case is as follows. First, since the strict invariance conditions of $X_1$ are solvable, one can select a control input at each vertex of $X_1$ satisfying the strict invariance conditions. At $x=0$, set $u=0$. Then, triangulate the point set consisting of the vertices of $X_1$ and the origin into $n$-dimensional simplices such that $0$ is a vertex in each simplex (always possible since $0\in X_1^{\circ}$). See Figure \ref{fig:rem_proof}. 
\begin{figure}[t]
\begin{center}
\includegraphics[scale=.22, trim = 10mm 150mm 10mm 60mm]{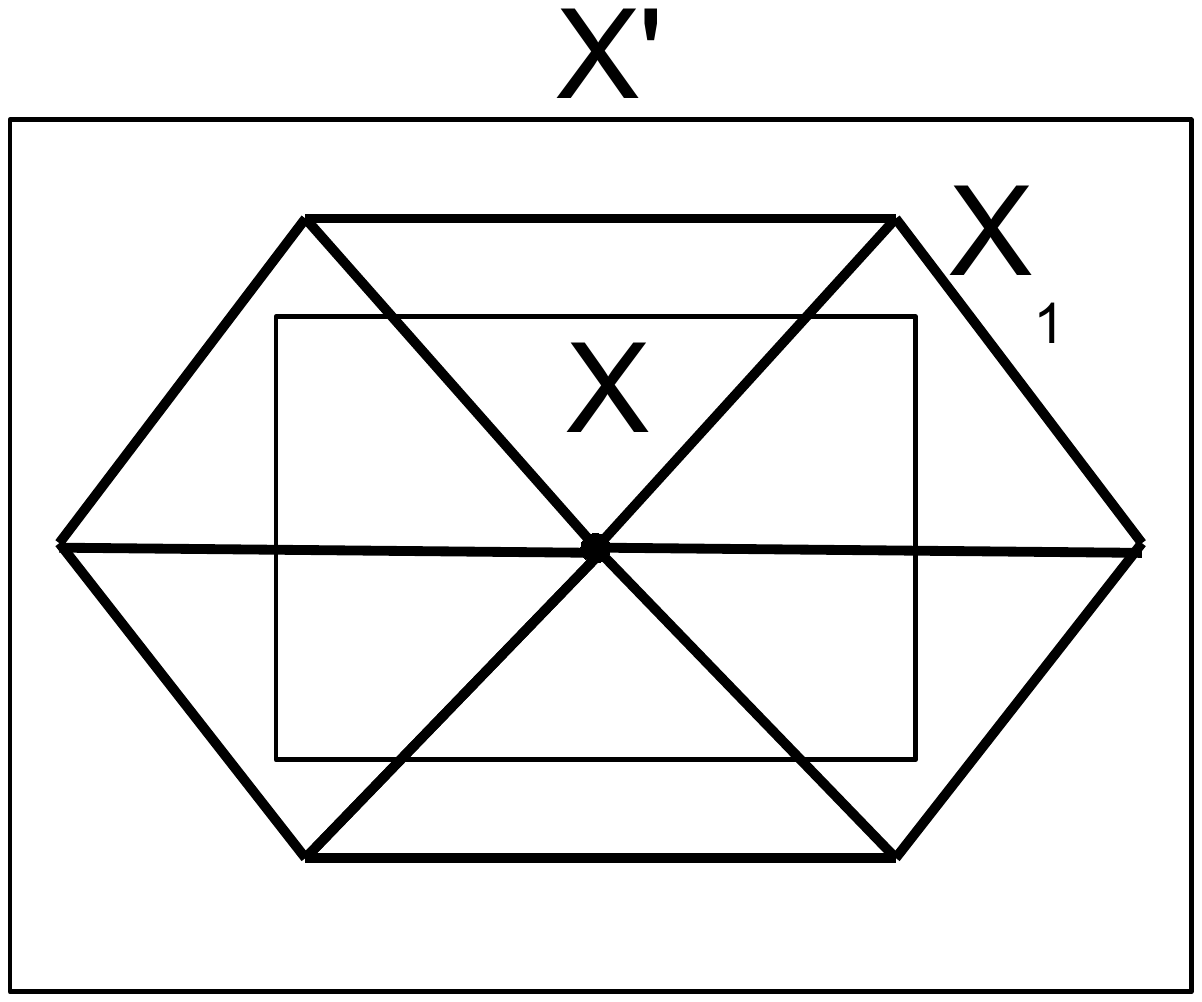} 
\end{center}
\caption{Illustrative figure for the triangulation of $X_1$ in Remark \ref{rem:suff}}
\label{fig:rem_proof}
\end{figure}
Using the control inputs selected at the vertices of the simplices of the triangulation, one can construct on each simplex a unique affine feedback such that the overall control law is a continuous piecewise affine (PWA) feedback satisfying the strict invariance conditions of $X_1$ \citep{HVS04}. It is straightforward to show this feedback stabilizes the origin. Starting from any $x\in X_1^{\circ}$, one can apply the obtained feedback to steer the system from $x$ to a point close to $0$ in finite time through $X_1^{\circ}$. Then, from this point, one can use the feedback in Theorem 5.8 of \citep{HC14_2} to reach $0$ in finite time through $X_1^{\circ}$. A similar proof can be provided for the backward dynamics to show that from the origin, the system can be steered to any $y\in X_2^{\circ}$ in finite time through $X_2^{\circ}$. Another condition under which the conditions (i)-(iii) of Theorem \ref{thm:caseB_main} are sufficient for RIBC is that $X_1$ and $X_2$ are positive invariant ellipsoidal sets associated with Lyapunov functions for the system \eqref{eq:thesystem2} and the backward dynamics, respectively. An example is given in Section \ref{sec:ex} (Example \ref{ex3_sec}). For the proof in this case, one can use the stabilizing feedback associated with the Lyapunov function instead of the continuous PWA feedback in the first part of this remark.           
\end{remark}

\begin{example}
\label{ex:caseB}
Consider again the system and the polytope in Example \ref{ex:mot2}. We have shown that system \eqref{sys:ex2} is not IBC w.r.t. $X$ in this example. Now it is required to check whether \eqref{sys:ex2} is RIBC w.r.t. $X$ through $X':=2.5X$. 

First, in this example, $\cO=\left\{x\in \RR^2~:~x_2=0\right\}$, the $x_1$-axis. 
Clearly, $X^{\circ}\cap \cO\neq \emptyset$ (Case B). We already verified in Example \ref{ex:mot2} that $(A,B)$ is controllable. Next, we check solvability of the invariance conditions of $X$. By solving an LP program at each vertex of $X$, we find that $u_1=1,~u_2=0,~u_3=-1$, and $u_4=0$ achieve the invariance conditions of $X$ at the vertices. As discussed before, this implies the invariance conditions of $X$ are solvable \citep{HVS04}. Therefore, condition (ii) of Theorem \ref{thm:caseB_main} is achieved with $X_1=X$. Then, as shown in Example \ref{ex:mot2}, the backward invariance conditions of $X$ are not solvable. Hence, we search for a simplicial polytope $X_2$ such that $X\subset X_2 \subseteq X'$ and the backward invariance conditions of $X_2$ are solvable. Here $\cB+\cO=\RR^2$. Following Remark \ref{rem:inv}, we find that $\bar{o}_2=(2.25,0)$ and $\bar{o}_4=(-2.25,0)$. Then, we define $X_2:=\conv\left\{v_1,\cdots,v_4,\bar{o}_2,\bar{o}_4\right\}$. By solving an LP at each vertex of $X_2$, one can verify that $u_1=0,~u_2=-4,~u_3=0,~u_4=4,~u_{\bar{o}_2}=0$, and $u_{\bar{o}_4}=0$ satisfy the backward invariance conditions of $X_2$. Thus, condition (iii) of Theorem \ref{thm:caseB_main} is achieved with $X_2=\conv\left\{v_1,\cdots,v_4,\bar{o}_2,\bar{o}_4\right\}$. We conclude from Theorem \ref{thm:caseB_main} that \eqref{sys:ex2} is RIBC w.r.t. $X$ through $X'=2.5X$. \demo    
\end{example}

\subsection{$X^{\circ}\cap \cO=\emptyset$ but $X'^{\circ}\cap \cO\neq \emptyset$}
This geometric situation is shown in Figure \ref{fig:O}(c). Since $X^{\circ}\cap \cO=\emptyset$, this geometric case is considerably more difficult. For instance, here it is not possible to simplify the problem to studying a linear system on a polytope $X$ satisfying $0\in X^{\circ}$, and so some simple proofs of the previous case do not work. Instead, using more involved proofs, we identify conditions for RIBC in this case. We start by showing that controllability of $(A,B)$ is a necessary condition for RIBC.
\begin{theorem}
\label{thm:caseC_1}
Consider the affine control system \eqref{eq:thesystem} and $n$-dimensional polytopes $X,~X'$ such that $X\subset X'$, $X^{\circ} \cap \cO=\emptyset$, and $X'^{\circ}\cap \cO\neq \emptyset$. If \eqref{eq:thesystem} is RIBC w.r.t. $X$ through $X'$, then $(A,B)$ is controllable. 
\end{theorem}     
\begin{pf}
Let $\bar{x}\in X'^{\circ}\cap \cO$, and let $\bar{u}$ be such that $A\bar{x}+B\bar{u}+a=0$. Define $\tilde{x}=x-\bar{x}$ and $\tilde{u}=u-\bar{u}$. Then, it is easy to verify that the dynamics in the new coordinates are $\dot{\tilde{x}}=A\tilde{x}+B\tilde{u}$. Let $X_n$, $X'_n$ represent the polytopes $X,~X'$ expressed in the new coordinates, respectively. It is clear that $0\notin X_n^{\circ}$ and $0\in X_n'^{\circ}$.
Now suppose by the way of contradiction that $(A,B)$ is not controllable. Let $P:=[i_1~\cdots~i_n]$, where $\left\{i_1,\cdots,i_k\right\}$ form a basis for the controllability subspace $\Im(Q_c)$ ($Q_c$ is the controllability matrix), and $\left\{i_1,\cdots,i_n\right\}$ form a basis for $\RR^n$. Then, let $z:=P^{-1}\tilde{x}$. The dynamics in the new coordinates are:
\begin{equation}
\label{Kalman_dec}
\left[ 
\begin{array}{rr}
\dot{z_1}\\\dot{z_2}\end{array} 
\right] =
\left[ 
\begin{array}{cc}
A_{11} & A_{12}  \\
0 & A_{22}  
\end{array} 
\right] 
\left[ 
\begin{array}{rr}
z_1\\ z_2\end{array} 
\right]
+
\left[ 
\begin{array}{rr}
B_1\\0\end{array} 
\right] u \,,
\end{equation} 
where $z_1\in \RR^k$, $z_2\in \RR^{n-k}$, $A_{11}\in \RR^{k\times k}$, $A_{12}\in \RR^{k\times (n-k)}$, $A_{22}\in \RR^{(n-k)\times (n-k)}$, and $B_1\in \RR^k$. Let $Z,~Z'$ denote the polytopes $X_n,~X'_n$ expressed in the new coordinates. Clearly, $0\notin Z^{\circ}$ and $0\in Z'^{\circ}$.    
By assumption, any two states $\bar{z}_1,~\bar{z}_2\in Z^{\circ}$ are mutually accessible through $Z'^{\circ}$. 
We study all the possible cases of the eigenvalues of $A_{22}$, and for each case, we reach a contradiction.
\begin{itemize}
\item [(i)] The subsystem $\dot{z_2}=A_{22}z_2$ is unstable: Let $\bar{z}_1,~\bar{z}_2\in Z^{\circ}$ be arbitrary. By assumption, there exist bounded control inputs that connect $\bar{z}_1$ to $\bar{z}_2$ in finite time through $Z'^{\circ}$. Also, there exist bounded control inputs that connect $\bar{z}_2$ to $\bar{z}_1$ in finite time through $Z'^{\circ}$. Therefore, starting at $\bar{z}_1\in Z^{\circ}$, the state trajectory $\phi(\bar{z}_1,t,u)$, $t\geq 0$, can be bounded by a proper selection of $u$. This can only happen if $\bar{z}_1$ has zero components in the directions of the eigenvectors associated with the unstable eigenvalues of $A_{22}$. But since $\bar{z}_1\in Z^{\circ}$ is arbitrary, then it must be that all points in $Z^{\circ}$ have zero components in the directions of the eigenvectors associated with the unstable eigenvalues of $A_{22}$, which clearly contradicts the fact that $\dim(Z)=n$.
\item [(ii)] The matrix $A_{22}$ has an eigenvalue $\lambda$ with a negative real part: Similar to the previous case, it can be shown that starting from any $\bar{z}_1\in Z^{\circ}$, the state trajectory of the backward dynamics can be bounded by a proper selection of $u$. But, this is impossible since the backward dynamics have an uncontrollable eigenvalue with a positive real part (an eigenvalue of the matrix $-A_{22}$), and so similar to the previous case, we reach a contradiction.
\item [(iii)] The matrix $A_{22}$ has an eigenvalue $\lambda=0$: By converting the dynamics $\dot{z_2}=A_{22}z_2$ to the Jordan form, it can be shown there exists $\beta \in \RR^n$ such that $\beta \cdot \dot{z}=0$ for any $z\in \RR^n$ and any $u\in \RR^m$. Therefore, starting at any $\bar{z}_1\in Z^{\circ}$, the state trajectory $\phi(\bar{z}_1,t,u)$ has a fixed $\beta$-component whatever control input is selected. Then, since $Z$ is an $n$-dimensional polytope, there exists $\bar{z}_2\in Z^{\circ}$ such that $\beta \cdot \bar{z}_1\neq \beta \cdot \bar{z}_2$. This implies $\bar{z}_2\in Z^{\circ}$ is not accessible from $\bar{z}_1\in Z^{\circ}$ whatever control input is selected, a contradiction.
\item [(iv)] All the eigenvalues of $A_{22}$ are complex with zero real part: For the repeated eigenvalues, we assume that the associated eigenvectors are linearly independent. For otherwise, the subsystem $\dot{z_2}=A_{22}z_2$ is unstable (Case (i) of the proof). Then, by a standard argument, there exist two perpendicular directions $\beta_1,~\beta_2$ such that $(\beta_1\cdot z(t))^2+(\beta_2\cdot z(t))^2=$constant, for all $t\geq 0$, where $z(t)$ denotes the state trajectory of \eqref{Kalman_dec} under some control input $u$ starting at a point $z\in \RR^n$. Let $\bar{z}_1\in Z^{\circ}$ be such that $\beta_1\cdot \bar{z}_1\neq 0$. This is always possible since $Z$ is an $n$-dimensional polytope. Also, since $\bar{z}_1\in Z^{\circ}$, there exists $\delta>0$ such that $B_{\delta}(\bar{z}_1)\subset Z^{\circ}$. Now, define $\bar{z}_2=\bar{z}_1+(\frac{\delta}{c})\beta_1$, where $c>0$ is selected sufficiently large such that $\bar{z}_2\in B_{\delta}(\bar{z}_1)\subset Z^{\circ}$, and the $\beta_1$-components of $\bar{z}_1$ and $\bar{z}_2$ have the same sign. Then, since $\beta_1,~\beta_2$ are perpendicular, $\beta_2\cdot \bar{z}_2=\beta_2 \cdot \bar{z}_1$. Thus, the value of $(\beta_1\cdot z)^2+(\beta_2\cdot z)^2$ evaluated at $\bar{z}_1$ is different from its value at $\bar{z}_2$. Hence, starting at $\bar{z}_1\in Z^{\circ}$, the state trajectory cannot reach $\bar{z}_2\in Z^{\circ}$, a contradiction.
\end{itemize} 
Since in all the above cases we reach a contradiction, we conclude that $(A,B)$ is controllable.
\end{pf}
Next, we present a second necessary condition for RIBC in this case. Recall that for system \eqref{eq:thesystem} on a compact convex set $X$, we say the invariance conditions of $X$ are solvable if for each $x\in \partial X$, there exists $u\in \RR^m$ such that $Ax+Bu+a \in T_{X}(x)$, and we say the backward invariance conditions of $X$ are solvable if for each $x\in \partial X$, there exists $u\in \RR^m$ such that $-Ax-Bu-a \in T_{X}(x)$ .   
\begin{theorem}
\label{thm:caseC_2}
Consider the affine control system \eqref{eq:thesystem} and $n$-dimensional polytopes $X,~X'$ such that $X\subset X'$, $X^{\circ} \cap \cO=\emptyset$, and $X'^{\circ}\cap \cO\neq \emptyset$. If \eqref{eq:thesystem} is RIBC w.r.t. $X$ through $X'$, then there exists an $n$-dimensional compact convex set $X_1$ such that (i) $X\subset X_1 \subseteq X'$, (ii) $X_1^{\circ}\cap \cO\neq \emptyset$, (iii) the invariance conditions of $X_1$ are solvable, and (iv) the backward invariance conditions of $X_1$ are solvable.
\end{theorem}
\begin{pf}
By assumption, there exists $M>0$ such that for every $x,~y\in X^{\circ}$, there exist $T_{xy}>0$ and a control input $u$ defined on $[0,T_{xy}]$ such that $\left\|u(t)\right\|\leq M$ and $\phi(x,t,u)\in X'^{\circ}$ for all $t\in [0,T_{xy}]$, and $\phi(x,T_{xy},u)=y$. Therefore, we have
\[
X^{\circ} \subseteq S:=\left\{\phi(x,t,u)~:~t\in[0,T_{xy}],~x,~y\in X^{\circ}\right\} \subseteq X'^{\circ}. 
\]
Since $X'^{\circ}$ is convex, then we have
\[
X^{\circ} \subseteq \conv(S) \subseteq X'^{\circ}. 
\]
Let $C:=\conv(S)$. First, we claim that $C^{\circ}\cap \cO\neq \emptyset$. For if not, then it can be shown using an argument similar to the proof of Theorem \ref{thm:caseA} that states of $X^{\circ}$ are not mutually accessible through $C$, a contradiction. 
Second, let $\bar{x}\in C^{\circ}\cap \cO$, and let $\bar{u}\in \RR^m$ be such that $A\bar{x}+B\bar{u}+a=0$.
This is always possible since $\bar{x}\in \cO$.
Define $\tilde{x}=x-\bar{x}$ and $\tilde{u}=u-\bar{u}$. The dynamics in the new coordinates are $\dot{\tilde{x}}=A\tilde{x}+B\tilde{u}$. Let $X_n,~X'_n,~S_n$, and $C_n$ denote the representations of the sets $X,~X',~S$, and $C$ in the new coordinates, respectively. Clearly, $0\in C_n^{\circ}$ and $0\notin X_n^{\circ}$. Also, we have $X_n^{\circ}\subset C_n \subseteq X_n'^{\circ}$. 

Next, we show there exists $M'>0$ such that for each $x_0\in C_n$, there exists a control input $u$ such that $\left\|u(t)\right\|\leq M'$ and $\phi(x_0,t,u)\in C_n$ for all $t\geq 0$. For $x_0\in S_n$, the state trajectory can remain in $C_n$ for all $t\geq 0$ by applying the uniformly bounded control inputs that connect $x_0$ to a point in $X_n^{\circ}$, say $y$, through $S_n$, and then from $y$ apply the uniformly bounded control inputs that connect $y$ to $x_0$ through $S_n$, and repeat this procedure for all future time. Instead, if $x_0\in C_n$ but $x_0\notin S_n$, then using an argument similar to the one used in the proof of Theorem \ref{thm:caseB_2}, there exists a control input $u$ such that $\left\|u(t)\right\|\leq M$ (where $M$ is as defined in Definition \ref{prob0_rel}) and $\phi(x_0,t,u)\in C_n$ for all $t\geq 0$. Using a similar argument, it can be also shown that there exists $M''>0$ such that for each $x_0\in C_n$, there exists a control input $u$ such that $\left\|u(t)\right\|\leq M''$ and the state trajectory of the backward dynamics $\dot{\tilde{x}}=-A\tilde{x}-B\tilde{u}$, denoted $\phi'(x_0,t,u)$, satisfies $\phi'(x_0,t,u)\in C_n$ for all $t\geq 0$.  

Now let $X_1:=\overline{C_n}$. Clearly, $X_1$ is convex and closed. Also, since $C_n\subseteq X_n'^{\circ}$, then $X_1\subseteq X'_n$, and so $X_1$ is bounded, hence compact. Moreover, $X_n \subset X_1$, and so $X_1$ is an $n$-dimensional set. Furthermore, $0\in C_n^{\circ}\subset X_1^{\circ}$, and so $X_1^{\circ}\cap \cO\neq \emptyset$. Next, we need to show that the invariance conditions of $X_1$ are solvable. Aided with the property proved in the previous paragraph, this follows from the proof of Theorem 4.3 and Remark 4.2 of \citep{HC14_2}. Similarly, the invariance conditions of $X_1$ w.r.t. the backward dynamics (the backward invariance conditions) can be proved. We conclude $X_1$ is an $n$-dimensional compact convex set satisfying the conditions (i)-(iv) of the theorem.          
\end{pf}
Now we present the main result for this geometric case.
\begin{theorem}
\label{thm:caseC_main}
Consider the affine control system \eqref{eq:thesystem} and $n$-dimensional polytopes $X,~X'$ such that $X\subset X'$, $X^{\circ} \cap \cO=\emptyset$, and $X'^{\circ}\cap \cO\neq \emptyset$. If \eqref{eq:thesystem} is RIBC w.r.t. $X$ through $X'$, then the following conditions hold:
\begin{itemize}
\item [(i)] $(A,B)$ is controllable.
\item [(ii)] There exists an $n$-dimensional convex compact set $X_1$ such that $X\subset X_1 \subseteq X'$, and the invariance conditions of $X_1$ are solvable.  
\item [(iii)] There exists an $n$-dimensional convex compact set $X_2$ such that $X\subset X_2 \subseteq X'$, and the backward invariance conditions of $X_2$ are solvable.   
\item [(iv)] The sets $X_1$ and $X_2$ in conditions (ii), (iii) satisfy $X_1^{\circ}\cap X_2^{\circ}\cap \cO\neq \emptyset$.    
\end{itemize}
Moreover, if $X_1,~X_2$ are simplicial polytopes, then the conditions (i)-(iv) are also sufficient for RIBC.
\end{theorem}
\begin{pf}
The necessity of (i) follows from Theorem \ref{thm:caseC_1}. Then, the necessity of (ii)-(iv) follows from Theorem \ref{thm:caseC_2} (Notice that in Theorem \ref{thm:caseC_2}, $X_2=X_1$ and $X_1^{\circ}\cap \cO\neq \emptyset$). 
Then, assume the conditions (i)-(iv) are achieved, and $X_1,~X_2$ are simplicial polytopes. We need to show \eqref{eq:thesystem} is RIBC w.r.t. $X$ through $X'$. Let $\bar{x}\in X_1^{\circ}\cap X_2^{\circ} \cap \cO$, and let $\bar{u}\in \RR^m$ be such that $A\bar{x}+B\bar{u}+a=0$. This is possible since $\bar{x}\in \cO$. Define $\tilde{x}=x-\bar{x}$ and $\tilde{u}=u-\bar{u}$. The dynamics in the new coordinates are $\dot{\tilde{x}}=A\tilde{x}+B\tilde{u}$. Let $X_n,~X'_n,~X_{1n}$, and $X_{2n}$ represent the sets $X,~X',~X_1$, and $X_2$ expressed in the new coordinates, respectively. Clearly, $0\in X_{1n}^{\circ} \cap X_{2n}^{\circ}$, $X_n\subset X_{1n} \subseteq X'_n$, and $X_n\subset X_{2n} \subseteq X'_n$. The rest of the proof is the same as the proof of the sufficiency part of Theorem \ref{thm:caseB_main}.  
\end{pf}
Similarly to Remark \ref{rem:suff}, it can be verified that conditions (i)-(iv) of Theorem \ref{thm:caseC_main} are also sufficient under the conditions mentioned in Remark \ref{rem:suff}.   
\begin{remark}
Notice that the conditions of Theorem \ref{thm:caseB_main} (for Case B: $X^{\circ}\cap \cO\neq \emptyset$) and Theorem \ref{thm:caseC_main} (for Case C: $X^{\circ}\cap \cO=\emptyset$ but $X'^{\circ}\cap \cO\neq \emptyset$) are almost the same. The difference is that in Theorem \ref{thm:caseC_main} we need to verify that $X_1,~X_2$ satisfy $X_1^{\circ}\cap X_2^{\circ}\cap \cO\neq \emptyset$, while in Theorem \ref{thm:caseB_main} this condition is automatically achieved since $X^{\circ}\subset(X_1^{\circ}\cap X_2^{\circ})$ and for the geometric case B, $X^{\circ}\cap \cO\neq \emptyset$.   
\end{remark}
\begin{example}
\label{ex:CaseC_2}
Consider the system
\begin{equation}
\label{sys:ex5}
\dot{x} =
\left[ 
\begin{array}{cc}
0 & 1  \\
0 & 0  
\end{array} 
\right] x +
\left[ 
\begin{array}{rr}
0\\1\end{array} 
\right] u  \,,
\end{equation}
and polytopes $X:=\conv\left\{v_1,\cdots,v_4\right\}$, $X':=\conv\left\{v_5,\cdots,v_8\right\}$ shown in Figure \ref{fig:ex5}, where $v_1=(0,0),~v_2=(1,0),~v_3=(1,1),~v_4=(0,1),~v_5=(-2,-1),~v_6=(2,-1),~v_7=(2,1)$ and $v_8=(-2,1)$. 
\begin{figure}[t]
\begin{center}
\includegraphics[scale=.24, trim = 10mm 110mm 10mm 40mm]{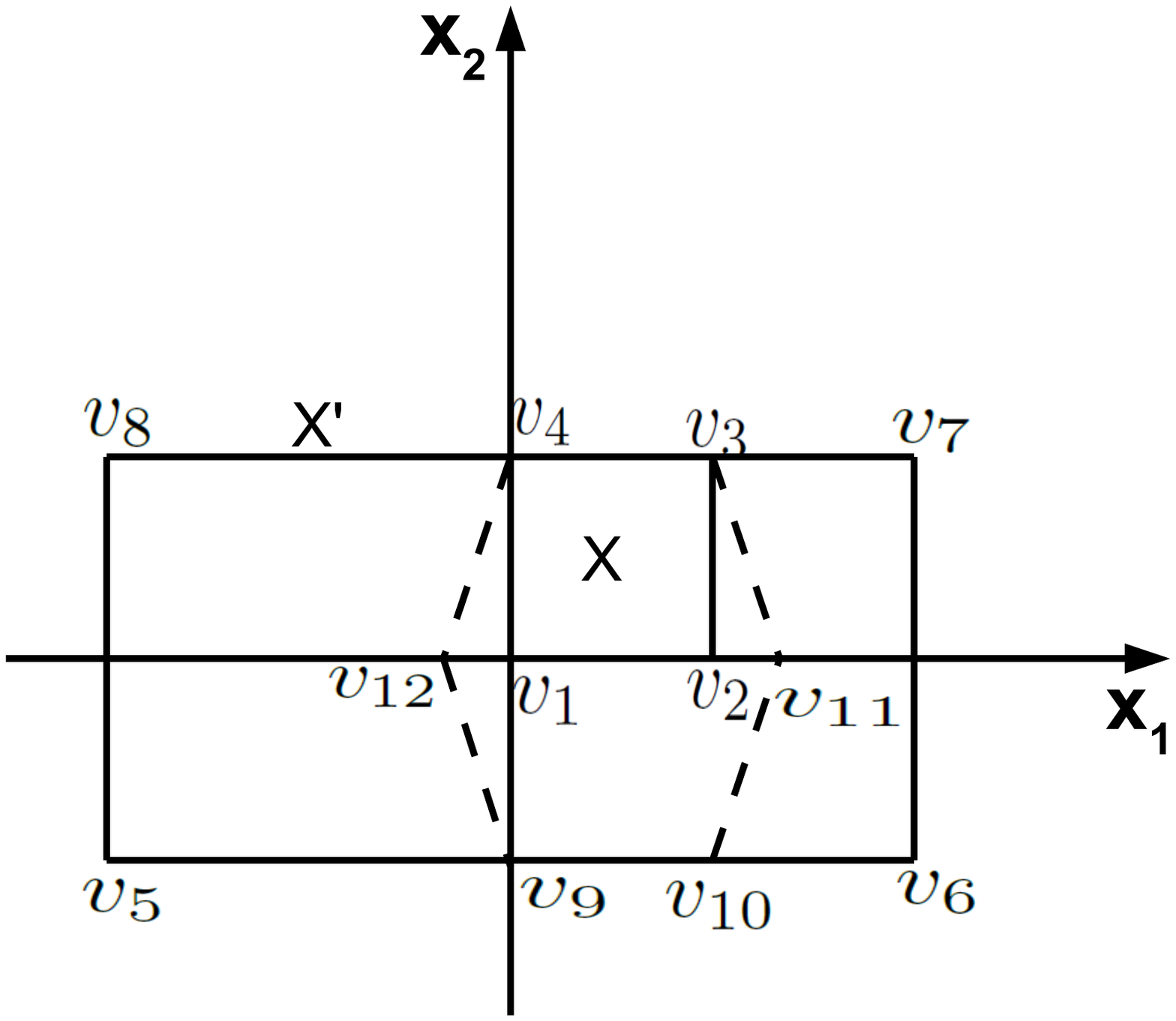} 
\end{center}
\caption{The polytopes $X$ and $X'$ in Example \ref{ex:CaseC_2}}
\label{fig:ex5}
\end{figure}
It is required to study whether \eqref{sys:ex5} is RIBC w.r.t. $X$ through $X'$. First, we calculate the set of possible equilibria $\cO$. We have $\cO=\left\{x\in \RR^2~:~x_2=0\right\}$, the $x_1$ axis. Thus, in this example $X^{\circ}\cap \cO=\emptyset$ but $X'^{\circ}\cap \cO\neq \emptyset$ (Case C). 
Next, it can be easily verified that $(A,B)$ is controllable. Now let $v_9:=(0,-1),~v_{10}:=(1,-1),~v_{11}:=(1.25,0)$, and $v_{12}:=(-0.25,0)$. Then, let $X'':=\conv\left\{v_3,v_4,v_9,v_{10},v_{11},v_{12}\right\}$. Clearly, $X''^{\circ}\cap \cO\neq \emptyset$. By solving an LP program at each vertex of $X''$, we find that the control inputs $u_3=-4,~u_4=0,~u_9=4,u_{10}=0,u_{11}=0$, and $u_{12}=0$ satisfy the invariance conditions of $X''$ at the vertices. Thus, the invariance conditions of $X''$ are solvable \citep{HVS04}. Similarly, it can be shown the backward invariance conditions of $X''$ are solvable. Therefore, the conditions (i)-(iv) of Theorem \ref{thm:caseC_main} are achieved (with $X_1=X_2=X''$). We conclude from Theorem \ref{thm:caseC_main} that \eqref{sys:ex5} is RIBC w.r.t. $X$ through $X'$ in this example. \demo
\end{example}
\section{Examples}
\label{sec:ex}
In this section, we provide three examples to show the motivations behind the relaxed IBC notion, and to clarify the main results of the paper.
\begin{example}
\label{ex1_sec}
Consider a cart moving on a bounded table. 
Let $x$ denote the position of the cart, $u$ denote the input force, $m$ denote the mass of the cart, and $b$ denote the coefficient of friction. The state space model of the system is:
\begin{equation}
\label{eq:ex1_sys}
\left[\begin{array}{cc}
\dot{x}_1\\ \dot{x}_2 \end{array}\right] =
\left[ 
\begin{array}{cc}
0 & 1  \\
0 & -\frac{b}{m}   
\end{array} 
\right] \left[\begin{array}{cc}
x_1\\ x_2 \end{array}\right] +
\left[ 
\begin{array}{rr}
0\\\frac{1}{m}\end{array} 
\right] u.
\end{equation}
Suppose that in this example, we have strict safety constraints: $-1\leq x_1\leq 1$, $-1\leq x_2\leq 1$ that should not be violated even in the transient period. These safety constraints define the polytope $X$ shown in Figure \ref{fig:sec_ex1_2}, where $v_1=(-1,-1)$, $v_2=(1,-1)$, $v_3=(1,1)$, and $v_4=(-1,1)$. 
\begin{figure}[t]
\begin{center}
\includegraphics[scale=.23, trim = 10mm 100mm 10mm 0mm]{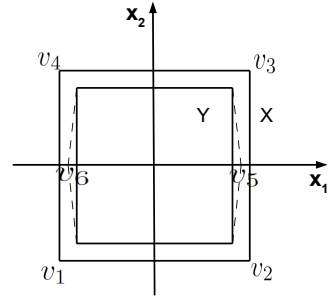} 
\end{center}
\caption{The polytopes $X$ and $Y$ in Example \ref{ex1_sec}}
\label{fig:sec_ex1_2}
\end{figure}
In the presence of these strict safety constraints, Kalman's controllability cannot be used to study mutual accessibility of states. Instead, we first study IBC of the system w.r.t. $X$. 
Although $(A,B)$ is controllable, at the vertex $v_3$, we have $Av_3+Bu_3=(1,-\frac{b}{m}+\frac{u_3}{m})$, and $h_1\cdot(Av_3+Bu_3)=1>0$, where $h_1=(1,0)$, whatever $u_3$ is selected. Thus, the invariance conditions of $X$ are not solvable at $v_3$, and from Theorem \ref{thm:main_IBC}, the system is not IBC w.r.t. $X$.\\
Next, we relax our control objective as follows. Suppose that the nominal operation of the system requires the soft constraints $-0.8\leq x_1\leq 0.8$ and $-0.8\leq x_2\leq 0.8$, resulting in the polytope $Y=\conv \{v_1',\cdots,v_4' \}$ shown in Figure \ref{fig:sec_ex1_2}, where $v_1'=(-0.8,-0.8)$, $v_2'=(0.8,-0.8)$, $v_3'=(0.8,0.8)$, and $v_4'=(-0.8,0.8)$. Starting from any initial position and speed in $Y^{\circ}$, can we reach any final position and speed in $Y^{\circ}$, without violating the strict safety constraints (through $X^{\circ}$)? To that end, we study relaxed IBC of the system w.r.t. $Y$ through $X$. In this example, we have $\cB=\spn\{(0,1)\}$ and $\cO=\left\{x\in \RR^2~:~x_2=0\right\}$. Hence, $0\in Y^{\circ}\cap \cO\neq \emptyset$ (Case B). Since $(A,B)$ is controllable, it remains to identify the sets $X_1,~X_2$ in Theorem \ref{thm:caseB_main} to show relaxed IBC. Notice that $\cO=\spn\{(1,0)\}$, and so $\cO+\cB=\RR^2$. By following the procedure in the first part of Remark \ref{rem:inv}, we construct the invariant polyope $X'=\conv \left\{v_1',\cdots,v_4',v_5,v_6\right\}$, where $v_5=(0.9,0)$ and $v_6=(-0.9,0)$. One can also verify that the backward invariance conditions of $X'$ are solvable. So, we have $X_1=X_2=X'$, and from Theorem \ref{thm:caseB_main}, the system is relaxed IBC w.r.t. $Y$ through $X$.  \demo  
\end{example}

\begin{example}
\label{ex2_sec}
We consider the mechanical system shown in Figure \ref{fig:sec_ex2}, in which we balance the center of mass above a pivot point.
\begin{figure}[t]
\begin{center}
\includegraphics[scale=.33, trim = 10mm 160mm 10mm 12mm]{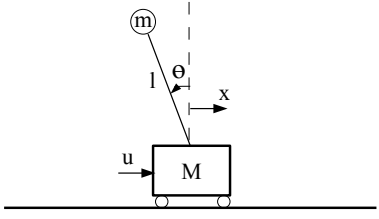} 
\end{center}
\caption{Illustrative figure for the balance system}
\label{fig:sec_ex2}
\end{figure} 
Examples of balance systems may include persons balancing sticks on their hands, humans standing upright, personal transporters, and rockets, among others \citep{Astrom}. Let $x_1$ and $x_3$ denote the position and velocity of the cart, respectively, while $x_2$ and $x_4$ denote the angle and angular rate of the structure above the cart, respectively. Assuming that $x_2$ and $x_4$ are close to zero, a linearized model of the system is \citep{Astrom} 
\begin{equation}
\label{eq:ex2_sys}
\left[\begin{array}{cccc}
\dot{x}_1\\ \dot{x}_2 \\ \dot{x}_3 \\ \dot{x}_4 \end{array}\right] =
\left[ 
\begin{array}{cccc}
0 & 0 & 1 & 0  \\
0 & 0 & 0 & 1 \\
0 & \frac{m^2l^2g}{\mu} & \frac{-cJ_t}{\mu} & \frac{-\gamma J_tlm}{\mu} \\
0 & \frac{M_tmgl}{\mu} & \frac{-clm}{\mu} & \frac{-\gamma M_t}{\mu}  
\end{array} 
\right] \left[\begin{array}{cccc}
x_1\\ x_2 \\ x_3 \\ x_4 \end{array}\right] + 
\left[ 
\begin{array}{rrrr}
0\\0 \\ \frac{J_t}{\mu}\\ \frac{lm}{\mu} \end{array} 
\right] u,
\end{equation}  
where $M$ is the mass of the cart, $J$ and $m$ are the moment of inertia and the mass of the system to be balanced, respectively, $l$ is the distance between the cart and the center of mass of the balanced body, $g$ is the gravitational acceleration constant ($9.8 m/{s^2}$), and $c$, $\gamma$ are coefficients of viscous friction. Also, $J_t=J+ml^2$ is the total inertia, $M_t=M+m$ is the total mass, and $\mu:=M_tJ_t-m^2l^2$. 
Now suppose that it is required to study whether we can mutually connect the states of the system having positive angle $x_2$, without changing the sign of the angle $x_2$. In particular, it is required to study mutual accessibility of the states in
\begin{equation*}
\begin{split}
X:=\big \{x\in \RR^4:-1\leq x_1\leq 1,0.2\leq x_2 \leq 0.3,\\ -0.1\leq x_3 \leq 0.1,-0.1\leq x_4\leq 0.1\big \}
\end{split}
\end{equation*}
through 
\begin{equation*}
\begin{split}
X':=\big \{x\in \RR^4:-1\leq x_1\leq 1,0\leq x_2 \leq 0.3,\\ -0.1\leq x_3 \leq 0.1,-0.1\leq x_4\leq 0.1\big \}.
\end{split}
\end{equation*}
To that end, we study relaxed IBC of the system w.r.t. $X$ through $X'$. First, we calculate the set of possible equilibria $\cO$. In this example, it can be verified that $\cO=\left\{x\in \RR^4~:~x_2=0,x_3=0,x_4=0\right\}$, the $x_1$-axis, and that $X'^{\circ}\cap \cO=\emptyset$ (Case A). From Theorem \ref{thm:caseA}, the system is not relaxed IBC w.r.t. $X$ through $X'$. To clarify more the obtained conclusion, let $\beta:=(0,-\frac{\gamma (M_tJ_t-J_tl^2m^2)}{\mu^2},\frac{lm}{\mu},-\frac{J_t}{\mu})$. It can be verified that $\beta \in \ker(B^T)$, i.e. $\beta \cdot B=0$, and $\beta \cdot (Ax)=(\frac{l^3m^3g}{\mu^2}-\frac{J_tM_tmgl}{\mu^2})x_2$. Since it is always the case that $M_tJ_t>l^2m^2$, then $\beta \cdot (Ax)\leq 0$, for all $x_2\geq 0$. Thus, starting at a point $x$, the $\beta$-component of the state trajectory is non-increasing as long as $x_2$ is non-negative, whatever $u$ is selected. In particular, let $x= (0,0.25,0,0.05)\in X^{\circ}$ and $y=(0,0.25,0,0)\in X^{\circ}$. We have $\beta \cdot x < \beta \cdot y$, and so starting from $x\in X^{\circ}$, we cannot reach $y\in X^{\circ}$ through $X'^{\circ}$. This clarifies the proof of Theorem \ref{thm:caseA}. \demo 
\end{example}    
\begin{example}
\label{ex3_sec}
Consider the linear circuit, shown in Figure \ref{fig:sec_ex3}, where $L_1=L_2=1H$, $C=1F$, and $R=1\Omega$. 
\begin{figure}[t]
\begin{center}
\includegraphics[scale=.22, trim = 10mm 40mm 10mm 30mm]{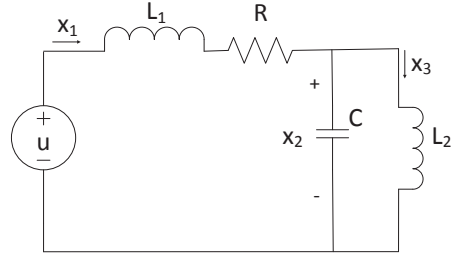} 
\end{center}
\caption{The linear circuit in Example \ref{ex3_sec}}
\label{fig:sec_ex3}
\end{figure}
The state space model of the system is
\begin{equation}
\label{eq:ex3_sys}
\left[\begin{array}{cccc}
\dot{x}_1\\ \dot{x}_2 \\ \dot{x}_3  \end{array}\right] =
\left[ 
\begin{array}{cccc}
-1 & -1 & 0   \\
1 & 0 & -1  \\
0 & 1 & 0 
\end{array} 
\right] \left[\begin{array}{cccc}
x_1\\ x_2 \\ x_3  \end{array}\right] + 
\left[ 
\begin{array}{rrrr}
1\\0 \\ 0 \end{array} 
\right] u.
\end{equation}
Let $X:=\left\{x\in \RR^3:-0.25\leq x_i\leq0.25,~i=1,\cdots,3\right\}$, and $X':=\left\{x\in \RR^3:-1\leq x_i\leq 1,~i=1,\cdots,3\right\}$. It is required to study mutual accessibility of the states of $X^{\circ}$ through $X'^{\circ}$. First, we calculate the set of possible equilibria $\cO=\left\{x\in \RR^3~:~x_1-x_3=0,~x_2=0\right\}$. We have $0\in X^{\circ}\cap \cO\neq \emptyset$ (Case B). Second, it is easy to verify that $(A,B)$ is controllable. Third, we identify a convex, compact set $X_1$ such that $X \subseteq X_1\subseteq X'$, and the invariance conditions of $X_1$ are solvable. Following Remark \ref{rem:inv}, we solve a set of LMIs to get a stabilizing feedback and a corresponding Lyapunov function. We find that 
$K_1~=~[-0.8587~-0.7274~0.1267]$, and the Lyapunov function is $V_1=x^TP_1x$, where
\[
P_1=
\left[
\begin{array}{cccc}
0.8587 & 0.7274 & -0.1267   \\
0.7274 & 2.3374 & 0.492  \\
-0.1267 & 0.492 & 2.1186 
\end{array}
\right].
\]
Let $X_1:=\left\{x\in \RR^3~:~x^TP_1x\leq 0.5\right\}$. It can be verified that the compact, convex set $X_1$ satisfies $X\subset X_1\subset X'$, and $u=K_1x$ satisfies the invariance conditions of $X_1$. Next, we identify a convex, compact set $X_2$ such that $X \subseteq X_2\subseteq X'$, and the backward invariance conditions of $X_2$ are solvable. Following Remark \ref{rem:inv}, we solve a set of LMIs to get a stabilizing feedback for the backward dynamics and a corresponding Lyapunov function. We find $K_2~=~[2.8587~-0.727~-0.1267]$, and the Lyapunov function for the backward dynamics is $V_2=x^TP_2x$, where
\[
P_2=
\left[
\begin{array}{cccc}
2.8587 & -0.7274 & -0.1267   \\
-0.7274 & 4.3374 & -0.492  \\
-0.1267 & -0.492 & 4.1186 
\end{array}
\right].
\]  
Let $X_2:=\left\{x\in \RR^3~:~x^TP_2x\leq 1\right\}$. It can be verified that the compact, convex set $X_2$ satisfies $X\subset X_2\subset X'$, and $u=K_2x$ satisfies the backward invariance conditions of $X_2$. Collecting all of the above together, we conclude from Theorem \ref{thm:caseB_main} that the system is relaxed IBC w.r.t. $X$ through $X'$.\\
For instance, suppose that it is required to connect $x_0=(0.2,0.2,0.2)\in X^{\circ}$ to $x_f=(-0.1,0.1,-0.1)\in X^{\circ}$ in finite time through $X'^{\circ}$. 
\begin{figure}[t] 
\begin{center}
\subfigure[]
{\includegraphics[trim = 0mm 70mm 0mm 65mm,clip,width = 0.8\linewidth]{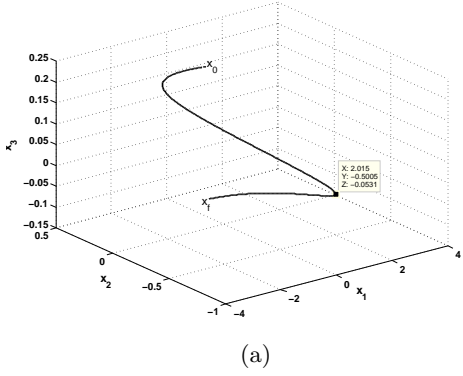}}
\subfigure[]
{\includegraphics[trim = 0mm 70mm 0mm 65mm,clip,width = 0.8\linewidth]{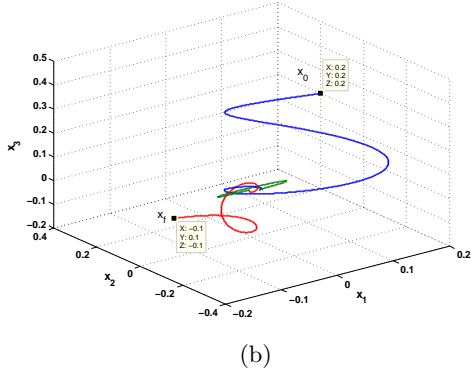}}
\caption{The state trajectories connecting $x_0$ to $x_f$ in Example \ref{ex3_sec} (a) under the traditional control law \eqref{eq:u}, (b) under the proposed control strategy}
\label{sim}
\end{center}
\end{figure}
Figure \ref{sim}(a) shows the trajectory obtained by applying the traditional control law \eqref{eq:u}, where $t_f=1~sec$. One can see that the trajectory reaches the point $(2.015,-0.5005,-0.0531)$, outside $X'$, which means it violates the strict safety constraints in this example. Instead, by following our proposed method in Remark \ref{rem:suff}, we first apply the control law $u=K_1x$ to steer the system to a point near the origin through $X_1^{\circ}\subset X'^{\circ}$ (the blue trajectory), then use the control law \eqref{eq:u} to connect the points close to the origin through $X'^{\circ}$ (the green trajectory), and finally apply $u=K_2x$ to steer the system to $x_f$ through $X_2^{\circ}\subset X'^{\circ}$ (the red trajectory). See Figure \ref{sim}(b). \demo 
\end{example}
\section{Conclusion}
\label{sec:con}
In this paper, we have extended the results of \citep{HC14_2,HC15_TAC} by studying the case where the given affine system is not in-block controllable with respect to the given polytope. In particular, we have introduced the notion of relaxed in-block controllability (RIBC) which studies mutual accessibility of the states in the interior of a given polytope through the interior of a given bigger polytope by applying uniformly bounded control inputs. By exploring all the possible geometric cases, we have provided necessary conditions for RIBC in all these cases. Moreover, we have shown when these conditions are also sufficient. Several examples have been provided to clarify the main results of the paper.
\section*{Appendix}
This appendix contains proofs included in this draft version for completeness.

\textbf{Proof of Lemma \ref{lem:tec1}:}\\
Let $x,~y\in X^{\circ}$ be arbitrary. Since $(A,B)$ is controllable, it is well known that for any $t_f>0$, $W_c(0,t_f):=\int _{0}^{t_f}e^{-A\tau}BB^Te^{-A^T\tau}d\tau$ is invertible, and the control input 
\begin{equation}
\label{eq:u}
u(t)=B^Te^{-A^Tt}W_c^{-1}(0,t_f)[-x+e^{-At_f}y],~t\in[0,t_f]
\end{equation}
steers the system from $x$ to $y$ in finite time $t_f$. Since $X$ is compact, $max_{z\in X}\|z\|$ exists, and so we can always identify a uniform upper bound $M>0$ such that $\left\|u(t)\right\|\leq M$ for all $t\in [0,t_f]$. By a straightforward argument, it can be shown that under \eqref{eq:u}, there exists a uniform $\lambda>1$ such that $\phi(x,t,u)\in (\lambda X)^{\circ}$ for all $t\in [0,t_f]$. 

\textbf{Proof of Theorem \ref{thm:caseA}:}\\
Since $X'^{\circ}\cap \cO=\emptyset$, then by Lemma 4.1 of \citep{HB13}, there exists $\beta \in \Ker(B^T)$, the perpendicular subspace to $\cB$, such that $\beta \cdot (Ax+Bu+a)<0$, for all $x\in X'^{\circ}$ and all $u\in \RR^m$. This implies for any $x\in X^{\circ}$ and under any control input, $\beta\cdot \phi(x,t,u)$ is non-increasing as long as $\phi(x,t,u)\in X'^{\circ}$. Then since $X$ is an n-dimensional polytope, we can identify two points $x,~y\in X^{\circ}$ such that $\beta \cdot x< \beta \cdot y$. Clearly, starting from $x$, the system cannot reach $y$ through $X'^{\circ}$.

\textbf{Sketch of the proof of Theorem \ref{thm:caseB_1}:}\\
The idea of the proof is that since $0\in X^{\circ}$, then for any $x,~y\in \RR^n$, there exists $c>0$ sufficiently large such that $\frac{x}{c},~\frac{y}{c}\in X^{\circ}$. Since $\frac{y}{c}$ is accessible from $\frac{x}{c}$ by assumption and \eqref{eq:thesystem2} is linear, then $y$ is accessible from $x$.  

\end{document}